\documentclass[fleqn,usenatbib]{mnras}

\usepackage{newtxtext,newtxmath}

\usepackage[T1]{fontenc}

\DeclareRobustCommand{\VAN}[3]{#2}
\let\VANthebibliography\thebibliography
\def\thebibliography{\DeclareRobustCommand{\VAN}[3]{##3}\VANthebibliography}

\usepackage{graphicx}	
\usepackage{amsmath}	
\usepackage{mathtools}
\usepackage{longtable}

\DeclarePairedDelimiterX{\infdivx}[2]{(}{)}{%
  #1\;\delimsize\|\;#2%
}

\DeclarePairedDelimiter{\norm}{\lVert}{\rVert}

\usepackage{color}
\definecolor{ao(english)}{rgb}{0., 0.5, 0.2}

\title[MICE-GC halo properties]{New catalogue of dark-matter halo properties identified in MICE-GC - I. Analysis of density profile distributions}

\author[Gonzalez et al.]{
Elizabeth J. Gonzalez,$^{1,2}$\thanks{E-mail: ejgonzalez@unc.edu.ar}
Kai Hoffmann,$^{3,4}$
Diego R. García Lambas,$^{1,2}$
Enrique Gazta\~{n}aga,$^{3,5}$
\newauthor
Dario Gra\~{n}a,$^{1}$
Pau Tallada-Cresp\'{i},$^{6,7}$
Jorge Carretero,$^{7,8}$
M. Victoria Santucho,$^{1,2}$
Pablo Fosalba,$^{3,5}$
\newauthor
Martin Crocce,$^{3,5}$
Francisco J. Castander,$^{3,5}$
Facundo Rodriguez,$^{1,2}$
and 
Martín Makler,$^{9,10}$
\\
$^{1}$ Instituto de Astronom\'{\i}a Te\'orica y Experimental (IATE-CONICET),
 Laprida 854, X5000BGR, C\'ordoba, Argentina.\\
$^{2}$ Observatorio Astron\'omico de C\'ordoba, Universidad Nacional de C\'ordoba, Laprida 854, X5000BGR, C\'ordoba, Argentina.\\
$^{3}$Institute of Space Sciences (ICE, CSIC), Campus UAB, Carrer de Can Magrans, s/n, 08193 Bellaterra (Barcelona), Spain\\
$^{4}$ Institute for Computational Science, University of Zurich, Winterthurerstr. 190, 8057 Zürich, Switzerland\\
$^{5}$ Institut d’Estudis Espacials de Catalunya (IEEC), Barcelona, Spain \\
$^{6}$ Centro de Investigaciones Energ\'{e}ticas, Medioambientales y Tecnol\'{o}gicas (CIEMAT), Avenida Complutense 40, 28040 Madrid, Spain \\
$^{7}$ Port d'Informaci\'{o} Cient\'{i}fica (PIC), Campus UAB, C. Albareda s/n, 08193 Bellaterra (Barcelona), Spain\\
$^{8}$ Institut de F\'{ı}sica d’Altes Energies (IFAE), The Barcelona Institute of Science and Technology, Campus UAB, 08193 Bellaterra (Barcelona), Spain \\
$^{9}$ International Center for Advanced Studies \& Instituto de Ciencias F\'isicas,  ECyT-UNSAM \& CONICET, 1650, Buenos Aires, Argentina\\
$^{10}$ Centro Brasileiro de Pesquisas F\'isicas, Rua Dr. Xavier Sigaud 150, CEP 22290-180, Rio de Janeiro, RJ, Brazil\\
}

\date{Accepted XXX. Received YYY; in original form ZZZ}

\pubyear{2015}

\begin{document}
\label{firstpage}
\pagerange{\pageref{firstpage}--\pageref{lastpage}}
\maketitle

\begin{abstract}
Constraints on dark matter halo masses from weak gravitational lensing can be improved significantly by using additional
information about the morphology of their density distribution, leading to tighter cosmological constraints derived from the halo mass function. This work is the first of two in which we investigate the accuracy of halo morphology and mass measurements in 2D and 3D. To this end, we determine several halo physical properties in the MICE-Grand Challenge dark matter only simulation. We present a public catalogue of these properties that includes density profiles and shape parameters measured in 2D and 3D, the halo centre at the peak of the 3D density distribution as well as the gravitational and kinetic energies and angular momentum vectors. The density profiles are computed using spherical and ellipsoidal radial bins, taking into account the halo shapes. We also provide halo concentrations and masses derived from fits to 2D and 3D density profiles using NFW and Einasto models for halos with more than $1000$ particles ($\gtrsim 3 \ 10^{13} h^{-1} M_{\odot}$). We find that the Einasto model provides better fits compared to NFW, regardless of the halo relaxation state and shape. The mass and concentration parameters of the 3D density profiles derived from fits to the 2D profiles are in general biased. Similar biases are obtained when constraining mass and concentrations using a weak-lensing stacking analysis. We show that these biases depend on the radial range and density profile model adopted in the fitting procedure, but not on the halo shape.
\end{abstract}




\section{Introduction}

The study of the halo density profiles is key for 
cosmology tests since, for example, accurate mass determinations can be used for establishing well-calibrated observational halo mass functions for comparison with cosmological predictions \citep{Bonamigo2015}. According to numerical simulations, stacked density distributions of dark matter particles of relaxed halos are well described across many orders of magnitudes in mass using simple density models \citep{Wang2020}. The most widely applied model to characterise the density distribution is the Navarro–Frenk–White \citep[NFW hereafter][]{Navarro1997} profile, which is characterised by two free parameters, a characteristic density and a scale radius. NFW profiles are usually set using mass and concentration parameters, where the mass is commonly defined as the one contained within a radius, $r_{200}$, that encloses a mean overdensity
of 200 times the critical density of the Universe. Besides, the concentration parameter is defined as the ratio between $r_{200}$ and the scale radius, which for halos of the same mass a higher value indicates that the object is denser in the central region. Simulations predict that concentrations are related to the halo mass, where the average halo concentrations weakly decrease with mass \citep[e.g. ][]{Navarro1997,Bullock2001,Maccio2007,Duffy2008,Bhattacharya2013,Okoli2017}.  

Halo NFW concentrations depend on the adopted radial range in which the fitting procedure is performed \citep{Gao2008}.
The differences obtained between fitted concentrations when varying the radial range are mainly caused by the deviations of the NFW profile from the observed density distributions, since the stacked profiles steepens more gradually with radius than this density model \citep{Navarro2004,Prada2006,Merritt2006}. The impact of the adopted fitted radial range can be reduced by using a density model that fits better the computed profiles, as the Einasto function \citep{Einasto1989,Retana-Montenegro2012}. This model is characterised by a logarithmic slope which is a power law of radius and is set in terms of an extra parameter.
The robustness of Einasto concentrations in terms of the adopted radial fitting range does not depend on this extra parameter, since similar concentration distributions are obtained regardless of whether it is fixed or fitted \citep{Gao2008,Klypin2016}.

 Although it is also well known that halos are not spherical but triaxial \citep[e.g.][]{Cole1996,Kasun2005,paz2006,Hoffmann2014,Vega-Ferrero2017,Gonzalez2021,Gonzalez2021b}, the mentioned models are based on the assumption of spherical symmetry. Nevertheless, they can be generalised using triaxial coordinates as proposed by \citet{Jing2002}. In a similar way, changing the radial coordinate of projected models is usually applied to consider the cluster shape when modelling surface density distributions, obtained using observational constraints as lensing properties \citep[e.g.][]{Adhikari2015,vanUitert2017,Okabe2020}. 
 
 Many physical aspects motivate the study of halo shapes since they are related with the formation processes of these structures. This include the study of the dark-matter particle nature \citep{Robertson2019}, intrinsic alignments \citep{Uitert2017b,Vedder2020} and assembly bias \citep{Lazeyras2017,Akitsu2021}.
 Moreover, the study of the interplay between halo shapes and density model parameters as masses and concentrations is necessary to calibrate mass-observable relations applied in cluster cosmology \citep{Zhang2022}. Even though that halo shapes have been studied in detail \citep[e.g.][]{Tenneti2014,Despali2017,Cataldi2020}, the connection of shape parameters with the density models has not been inspected in depth. At the same time, it is also necessary to take into account projected quantities, which can be useful to provide observational constraints and to estimate potential biasses. 
 Taking this into account, we aim to study the relation between halo shape parameters and several halo properties and how they can be tested through observational studies.
To this end, we present a series of papers devoted to the study of the relation between the halo shapes, environmental properties and relaxation state and how these parameters can be determined using observational data-sets.

Here we present a catalogue of halo properties based in the dark-matter only Marenostrum Institut de Ciències de l’Espai Grand Challenge (MICE-GC) simulation\footnote{\href{http://maia.ice.cat/mice/}{http://maia.ice.cat/mice/}} \citep{Fosalba2015a,Fosalba2015b,Carretero2015,Crocce2015}. The catalogue released details the main properties of the halos as their semi-axis obtained using the shape tensor, density profiles and kinetic and gravitational energies. Density profiles are computed in 3D as well as in projection using spherical and ellipsoidal binning. Moreover, for cluster-sized halos (those with more than 1000 particles) we provide the parameters that best describe the density distributions according to NFW and Einasto models. The large size-box of MICE-GC ($\sim 3\,h^{-1}$Gpc) gives an important statistical power to inspect the halo properties through a large mass and redshift range specially for higher mass halos. 

In this 
paper, the first of the series, we define the parameters that characterise the halo properties provided in the catalogue and analyse in depth the fitted density profiles of the selected cluster-sized halos and their relation with other physical parameters such as their shapes and masses. We analyse the fitting performance of the mentioned models and compare the derived masses with the total halo mass according to the linked dark-matter particles (FOF mass). We study the differences between the fitted parameters using ellipsoidal binning to compute the density profiles instead of the usual spherical binning. Furthermore, we compare fitted 3D parameters with those obtained using projected density distributions to evaluate the potential differences with observational constraints. Finally, we analyse the improvement of lensing studies with the new parameters provided, which are useful to test how shape parameters can be derived from observational data-sets and also to asses how mass estimates are affected by the assumed spherical shapes. The work is organised as follow, we describe the simulated data and define the computed halo properties in Sec. \ref{sec:prop}. Details of the columns of the catalogue presented  are available in the Appendix \ref{app:catalog}. In Sec. \ref{sec:analysis} we present the analysis of the fitted parameters from the 
density distributions and their relation with halo shapes. Finally we summarise and conclude in Sec. \ref{sec:summary}.


\section{Halo properties}
\label{sec:prop}

In this work, we use version 2 MICE-GC simulation developed using \textsc{gadget-2} \citep{Springel2005}. The simulation includes $4096^3$ dark matter particles within a box size of $3072\,h^{-1}$Mpc. The adopted cosmological model assumes $\Omega_m = 0.25$, $\Omega_\Lambda = 0.75 $, $\Omega_b = 0.044 $, $n_s = 0.95$, $\sigma_8 = 0.8$ and $h = 0.7$ and the softening length is set at $50\,h^{-1}$\,kpc. This results in a mass particle resolution of $m_p = 2.93 \times 10^{10} h^{-1} M_\odot$.

For the analysis, we run all over the MICE halo catalogue presented in \citet{Crocce2015} which is generated using a public halo finder\footnote{\href{http://www-hpcc.astro.washington.edu/}{http://www-hpcc.astro.washington.edu/}}.
Halos were identified using a friends-of-friends (FOF) algorithm with a linking length of $b = 0.2$ times the mean inter-particle distance. Only halos with more than ten particles are taken into account, which corresponds to a lower mass limit of $M_\text{FOF} \sim 2.2 \times 10^{11} M_\odot$. 
In the next subsections, we characterise the physical parameters obtained for the halos. Throughout this work all distances are expressed in physical units. 

\subsection{Relaxation indicators}
\label{subsec:relaxind}
\begin{figure*}
    \includegraphics[scale=0.55]{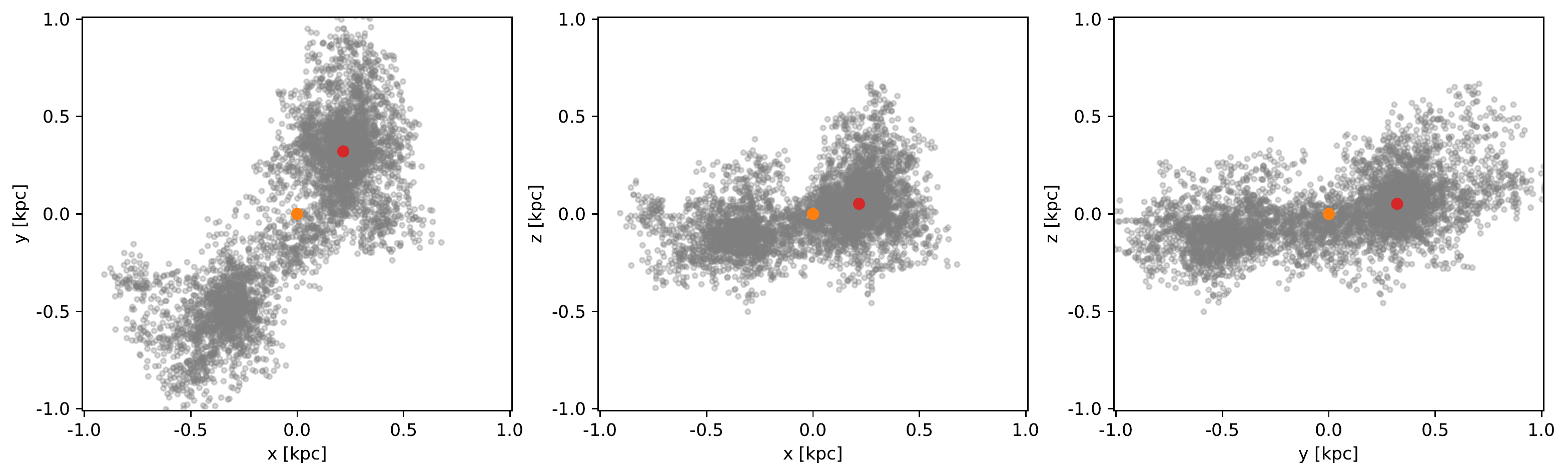}
    \caption{Dark matter particle distribution coordinates of the main axis of the simulation (grey dots) for a randomly selected non-relax halo, centred according to the coordinates given by the FOF identification (orange dot). The red dot corresponds to the location of the centre obtained from the shrinking-sphere method.}
    \label{fig:recenter}
\end{figure*}

Taking into account that dark-matter halos are evolving structures, several indicators can be derived in order to estimate the relaxation state of these systems. We derive quantities that can be related to two commonly used indicators \citep[e.g.,][]{Gao2008,Meneghetti2014,Despali2017}: the centre of mass displacement and the virial ratio. We compute the kinetic and gravitational self-potential energies defined as:
\begin{equation}
    K = \frac{1}{2} m_p \sum_{j=1}^{N_p} \norm{\vec{v}_j}^2,\,\,\,\, W = - G m^2_p  \sum_{j=1}^{N_p-1} \sum_{k=j+1}^{N_p} \norm{\vec{r}_{jk}}^{-1}
\end{equation}
where $\vec{v}_j$ is the velocity of the $j-$th particle with respect to the FOF centre of mass and $\vec{r}_{jk}$ is the distance vector between the $j-$th and $k-$th particles and the sum runs over the total number of particles identified for each halo, $N_p$. In this case, the energy virial ratio $2K/|W|$ indicates the relaxation halo state, where halos with $2K/|W| < 1.35$ are commonly defined as relaxed halos.

In order to provide an additional centre definition that is less affected by 
substructures, we recompute the halo centre using an iterative method based on a shrinking-sphere algorithm, similarly as the one described in \citet{Power2003}. This centre is obtained as the centre of mass according to the location of the particles within a shrunken sphere. This is performed by iteratively reducing the sphere radius by ten per cent in each iteration and recentring the sphere in each step at the last computed barycentre. The iteration stops either at the $10^\text{th}$ step or when the number of particles within the shrunken sphere is lower than 10. After applying this technique, the resultant centre is located in the densest substructure. In Fig. \ref{fig:recenter} we show the 2D particle distribution in the main axis of the simulation, for a halo with the FOF centre of mass and the shrinking-sphere centre location. As it can be noticed for this particular halo, the new computed centre is relocated at the densest region. This procedure is applied to the whole sample of halos. 

The scaled offsets, defined as $r_{c}/r_\text{max}$ (where $r_{c}$ is the distance from the FOF centre of mass to the shrink-sphere centre and $r_\text{max}$ is the radius of the first sphere which encloses all the FOF particles), could be used as a dynamical state indicator of the halo since more relaxed halos are expected to have lower offsets \citep{Bonamigo2015}. Hence, we define as relaxed halos in this work as those with $r_{c}/r_\text{max} < 0.1$ and $2K/|W| < 1.35$, which are commonly adopted values used to select dynamically relaxed systems in the literature \citep[e.g.][]{Neto2007,Skibba2011,Ludlow2013,Klypin2016,Brinckmann2018}.

\subsection{Main semi-axis shapes}

The semi-axis of the dark matter distribution for each halo are obtained according to the eigenvectors and eigenvalues of the moment of inertia tensor.
We obtain these quantities for the 3D dark matter particle distribution as well as for the 2D distribution that can be related to observable estimates. Projected shape parameters are computed by defining for each halo a tangential plane perpendicular to the line-of-sight vector, pointing to the halo centre. We use two common definitions of the shape tensor, a \textit{standard} tensor defined as
\begin{equation}
I_{ij} \equiv  \frac{\sum_{n}^{N_p}  x_{i n} \, x_{j n}}{N_p}
\end{equation}
and the \textit{reduced} tensor
\begin{equation} \label{eq:ired}
I^{r}_{ij} \equiv   \frac{ \sum_{n}^{N_p} (x_{i n} \, x_{jn}) / r_{n}^2 }{N_p},
\end{equation}
where $x_{in}$, $x_{jn}$ are the coordinates of the vector position of the $n^{th}$ dark matter particle with respect to centre computed using the shrinking-sphere algorithm (see \ref{subsec:relaxind}).
The sub-indexes $i,j$ are between $i,j\in[0,1]$ and $i,j\in[0,1,2]$ for 2D and 3D coordinates, respectively, and $r_n$ is the distance to the halo shrink-sphere centre.

By diagonalising the defined tensors we obtain the semi-axis modules
($A > B > C$ in 3D and $a > b$ in 2D) according the square-root of the eigenvalues, while the eigenvectors define the main axis directions. We obtain \textit{standard} and \textit{reduced} quantities in 3D as well as in projection. From these quantities we can obtain the usual shape parameters, the 3D sphericity $S = C/A$ and the 2D semi-axis ratio $q = b/a$, where $S \rightarrow 1$ and $q \rightarrow 1$ describe rounder shapes. Also, we can obtain the 3D triaxility defined as:
\begin{equation}
    T = \frac{1 - (B/A)^2}{1 - (C/A)^2},
\end{equation}
where $T \rightarrow 1$ indicate a more prolate distribution of particles. We obtain these parameters for the \textit{standard} and the \textit{reduced} tensors, where the \textit{reduced} quantities are specified with a subindex $r$ ($T_r$, $S_r$ and $q_r$).

\begin{figure}
    \includegraphics[scale=0.6]{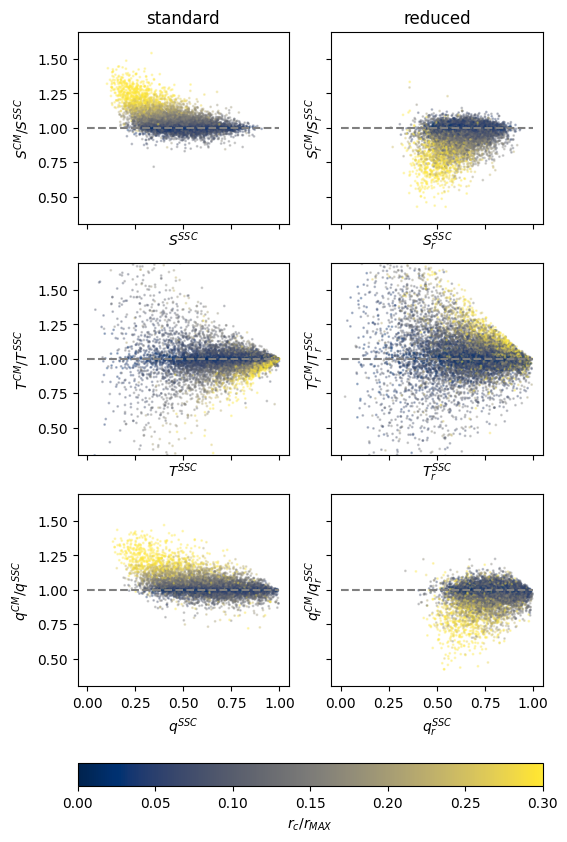}
    \caption{Shape parameters determined adopting the shrink-sphere centre ($SSC$ supra-index) compared to those obtained according to the FOF centre of mass ($CM$ supra-index). The colour code correspond to the scaled offset. Standard and reduced parameters are plotted in the left and right columns, respectively. Grey dashed line correspond to a one-to-one relation.}
    \label{fig:shape_off}
\end{figure}

Shape parameters are sensitive to the centre definition, since they are determined using the particle coordinates with respect to the adopted centre. In Fig. \ref{fig:shape_off} we show the shape parameters obtained for a random subsample of halos computed according to the FOF and the shrink-sphere centre. It can be noticed that the centre definition affects shape estimates differently depending on the use of standard or inertial tensors. Standard shapes of halos with higher scaled offsets values determined according to the FOF centre tend to predict rounder shapes. This is the result of a more spherically symmetric particle distribution with respect to the FOF centre of mass. On the other hand, reduced shapes are rounder when using the shrink-sphere centre definition. Taking into account that reduced quantities are obtained by up-weighting the distance centre, they also show a higher scatter from the one-to-one relation. In the rest of the analysis, we adopt the shrink-sphere centre to compute shape parameters since they are less affected by external substructure and are more related with the densest halo region. Moreover this centre is related with the peak of the density distribution which is commonly set for the density profile analysis. However, we provide in the released catalogue both shape parameters computed using the two centre definitions.

\subsection{Computation and fitting details}

\subsubsection{Profile modelling}

Radial density distributions can be well described according to different models. In particular, we use NFW and Einasto density models to describe the density distribution of the analysed halos. The NFW model is widely used to characterise the density distribution of dark matter halos and is given by
\begin{equation} \label{eq:nfw}
\rho(r) =  \dfrac{\rho_{\rm crit} \delta_{c}}{(r/r_{s})(1+r/r_{s})^{2}},
\end{equation}
where  $r_{s}$ is the scale radius and $\rho_{\rm crit}$ is the critical density of the Universe at the halo redshift. $\delta_{c}$ is the cha\-rac\-te\-ris\-tic overdensity
\begin{equation}
\delta_{c} = \frac{\Delta}{3} \dfrac{c_{\Delta}^{3}}{\ln(1+c_{\Delta})-c_{\Delta}/(1+c_{\Delta})},
\end{equation}
that depends on the halo concentration defined as $c_{\Delta} = r_{\Delta}/r_s$ which is the ratio size of the halo. $r_{\Delta}$ is the radius that encloses a mean overdensity of $\Delta \times \rho_{\rm crit}$. In this study we fix $\Delta = 200$ and we obtain the mass \mbox{$M_{200}=200\,\rho_{\rm crit} (4/3) \pi\,r_{200}^{3}$} as the mass enclosed within $r_{200}$. Thus, the profile can be parametrised by only two parameters $M_{200}$ and $c_{200}$. 

Einasto density model is defined as
\begin{equation} \label{eq:ein}
\rho(r) = \rho_{-2} \exp{\left( -\frac{2}{\alpha} \left[ \left( \frac{r}{r_{-2}} \right) ^\alpha  - 1 \right] \right)},
\end{equation}
where $\rho_{-2}$ sets the amplitude of the profile and is the density at $r = r_{-2}$, i.e., at the radius where the logarithmic slope of the density profile is $-2$ and which is the equivalent to the NFW scale-length, $r_s$. This model is characterised by a logarithmic slope which is a power law of radius and it is set in terms of the extra parameter, $\alpha$. 
In this case, we characterise the concentration of the profile according to $c_{200} = r_{200}/r_{-2}$. 
Surface density distributions for both models can be obtained by integrating equations \ref{eq:nfw} and \ref{eq:ein} along the line of sight, where the resultant equations that describe the projected distributions for NFW and Einasto models can be found in \citet{Wright2000} and \citet{Retana-Montenegro2012}, respectively.

According to previous studies, the Einasto model provides a more precise description of the radial density distribution in dark matter halos, which is also more robust against the variations in fitting details as, for example, the radial range used to fit the parameters \citep{Gao2008,Meneghetti2014}. Since halo density profile resolution varies with mass, where less-massive halos are more poorly resolved, force fitting NFW model can induce spurious correlations between mass and concentration \citep{Klypin2016}. Thus, adopting a better modelling as Einasto could provide less biased determinations. The good performance of the Einasto profile in modelling the density distribution is achieved even when the extra parameter, $\alpha$, is fixed
according to halo mass \citep{Gao2008}. Nevertheless, evaluating the concentration is not as straightforward as for the NFW model, where it is uniquely defined and halos with a denser centre are described with higher concentration values. Yet, for the Einasto model, $\alpha$ impacts not only in the shape but also in the concentration, thus the halo concentration is indeed properly described by $r_{200}/r_{-2}$ and $\alpha$ parameters together \citep{Klypin2016}. 

\subsubsection{Computation and parametrization of density profiles}
\label{subsec:fitdens}
We compute the halo density profiles by measuring the halo mass within 25 equally-spaced spherical and elliptical radial bins from 10 kpc$/h$ up to 1 Mpc$/h$. The density is computed up to a radial distance defined as the maximum value between $220\,\text{kpc}/h$ and $0.7 r_\text{max}$. In order to compute ellipsoidal profiles we first rotate halo particles according to the main axis of the particle distribution, defined in the previous subsection.
Ellipsoidal profiles in 3D are obtained by considering ellipsoidal shells according to a radial distance defined as $r^3 = \mathsf{ABC}$, which is computed assuming a constant semi-axis ratio across the whole radial range taking into account the \textit{standard} semi-axis modules defined in the previous subsection (i.e. $B/A$ and $C/A$ constant). In an analogous way, we obtain 2D elliptical profiles taking into account elliptical rings within a radial distance defined as $R^2 = \mathsf{ab}$, but in this case we consider $q$ as constant. Thus, for each halo, four different profiles are obtained, according to the spherical and ellipsoidal distribution, both in 3D, $\rho_r$, as well as in projection, $\Sigma_r$. 

Derived profiles are fitted using the publicly available Colossus\footnote{\href{https://bitbucket.org/bdiemer/colossus/src/master/}{https://bitbucket.org/bdiemer/colossus/src/master/}} astrophysics toolkit \citep{Diemer2018} using least-squares minimisation method. We adopt shot-noise errors which are computed in each bin as the mass particle resolution scaled by the shell volume and the ring area for 3D and 2D profiles, respectively. We fit individual profiles for halos with more than 1000 particles and more than 5 radial bins, from 50 kpc$/h$ taking into account the softening length.
From the fitting procedure, we obtain the defined masses, $M_{200}$, and concentrations, $c_{200}$ for both models introduced in the last section, and $\alpha$ in the case of Einasto. It is worth noting that in this particular work we fit the models for the individual halos instead of a stacked profile as usually done. Therefore, the presence of substructure in individual halos that could bias the fitted parameters must be taken into account in order to evaluate the results. However, this analysis allows us to study individually the impact of the relaxation state estimators in the fitted parameters. 

We evaluate how well profiles are described by the adopted models according to the residuals defined as in \citet{Meneghetti2014} in 3D as:
\begin{equation} \label{eq:res3d}
    \delta_{3D} = \frac{1}{N_\text{doff}} \sum_i [\log \rho_i - \log \rho(r_i,\textbf{p})]^2,
\end{equation}
where $N_\text{doff}$ is the number of degree of freedom (i.e. the number of fitted bins minus the number of free parameters), $\textbf{p}$ is the vector that contains the fitted parameters and the sums runs over the considered radial bins. In analogous way, residuals of 2D density distributions are obtained as: 
\begin{equation} \label{eq:res2d}
    \delta_{2D} = \frac{1}{N_\text{doff}} \sum_i [\log \Sigma_i - \log \Sigma(r_i,\textbf{p})]^2.
\end{equation}


\begin{figure}
    \includegraphics[scale=0.6]{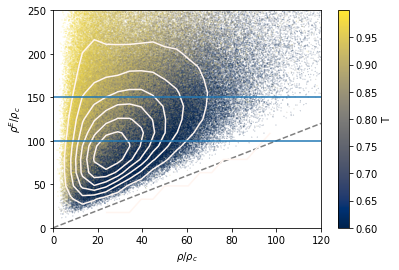}
    \caption{Ellipsoidal versus spherical halo three-dimensional densities scaled according to the critical density. Colour code corresponds to the halo triaxiality computed using the standard shape tensor, where higher values are related with a more prolate dark matter particle distribution. White contours trace the density levels of the data points. The grey dashed line corresponds to the one-to-one relation.}
    \label{fig:density}
\end{figure}

\section{Analysis of the halo properties}
\label{sec:analysis}

In this section, we present the analysis of the halo properties characterised through the estimated parameters presented above. Although shape, angular momentum vectors and relaxation indicators are obtained and presented in the released catalogue for all the halos, here we restrict our study to the halos that have fitted density profiles, thus more than 1000 particles ($M_\text{FOF} \gtrsim 2.9\times10^{13}$). We also consider all the halos with fitted mass and concentrations for the four profiles in the ranges of $1 < c_{200} < 15$ and $ \log M_{200} > 12$, taking into account both Einasto and NFW models. This reduce the total sample of fitted halos in a $35 \%$. In the case of relaxed halos, $92\%$ and $83\%$ of the halos have fitted Einasto and NFW parameters, respectively, within the adopted ranges. This percentages are higher if we consider only the fitted 3D spherical distributions, increasing up to $96\%$ and $99\%$, respectively. The presented analysis does not intend to address all the observed biases when comparing different fitting methodologies as well as projection effects, but mainly to validate the presented parameters and provide worthy information for the catalogue usage.

\subsection{Fitting performance of density distribution models}
\label{subsec:fit}
\begin{figure}
    \includegraphics[scale=0.6]{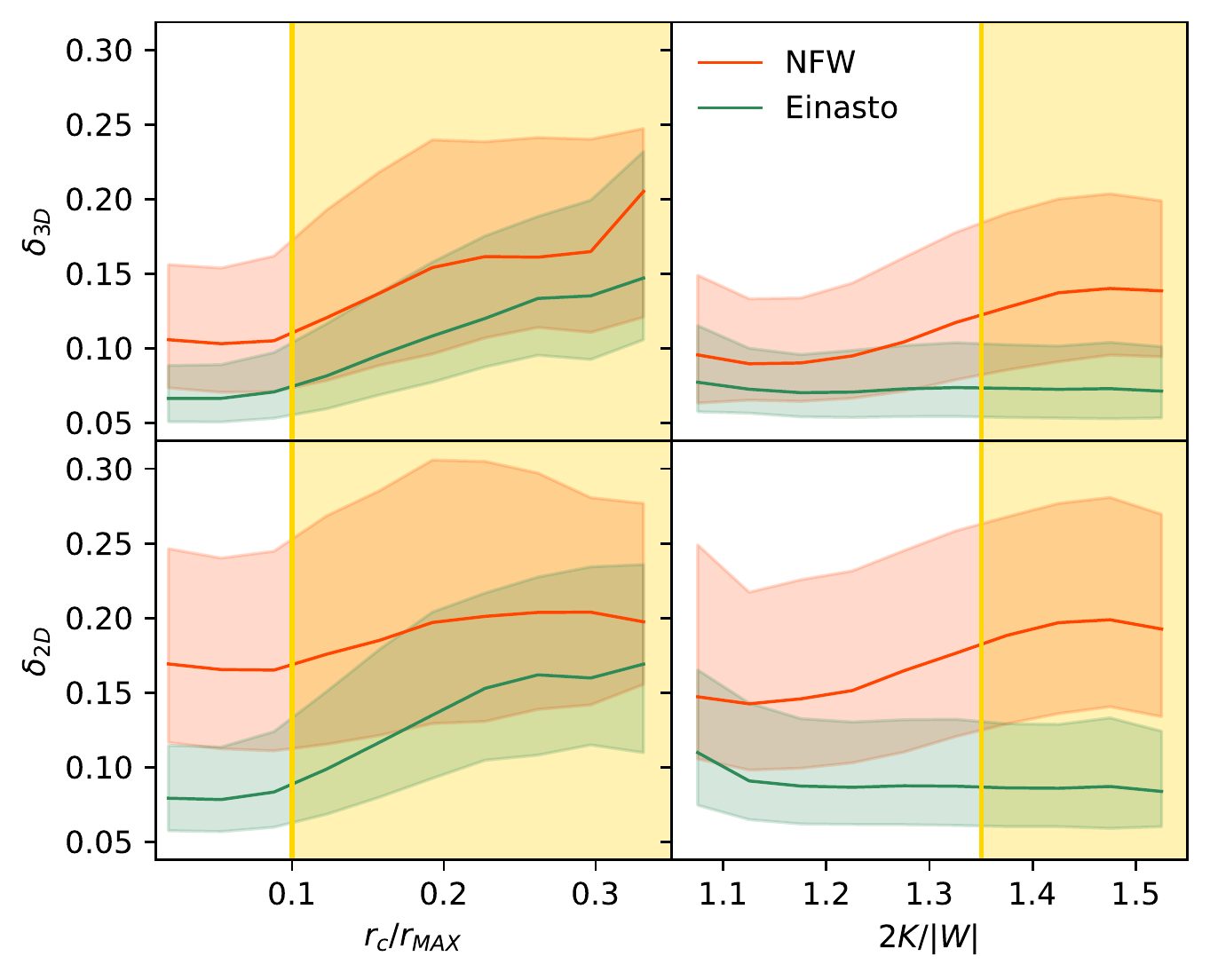}
    \caption{Median 3D and 2D residuals obtained in offsets and virial energy ratios by fitting NFW (orange line) and Einasto profiles (green line), for halos with ellipsoidal densities $100 < \rho^E/\rho_c < 150$. Shadow regions enclose from 25th up to 75th of the residual distributions. Yellow vertical lines are related with the cuts adopted to select relaxed halos and the shadow yellow region include the non-relaxed halos. }
    \label{fig:res}
\end{figure}

\begin{figure}
    \includegraphics[scale=0.6]{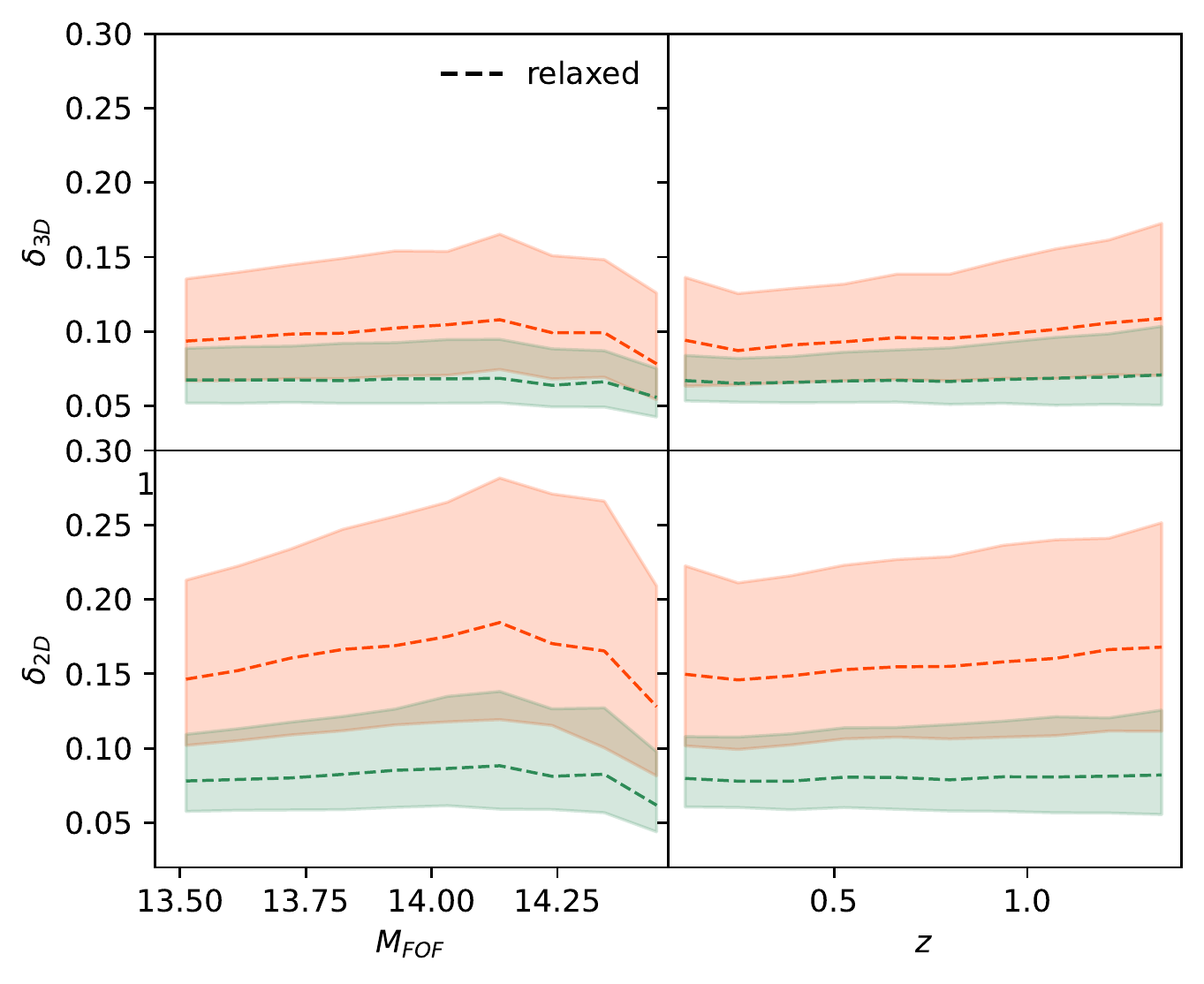}
    \caption{Median 3D and 2D residuals obtained in FOF masses and redshift bins, by fitting NFW (orange line) and Einasto profiles (green line), for only relaxed halos with ellipsoidal densities $100 < \rho^E/\rho_c < 150$. Shadow regions enclose from 25th up to 75th of the residual distributions.  }
    \label{fig:res_relax}
\end{figure}

\begin{figure}
    \includegraphics[scale=0.6]{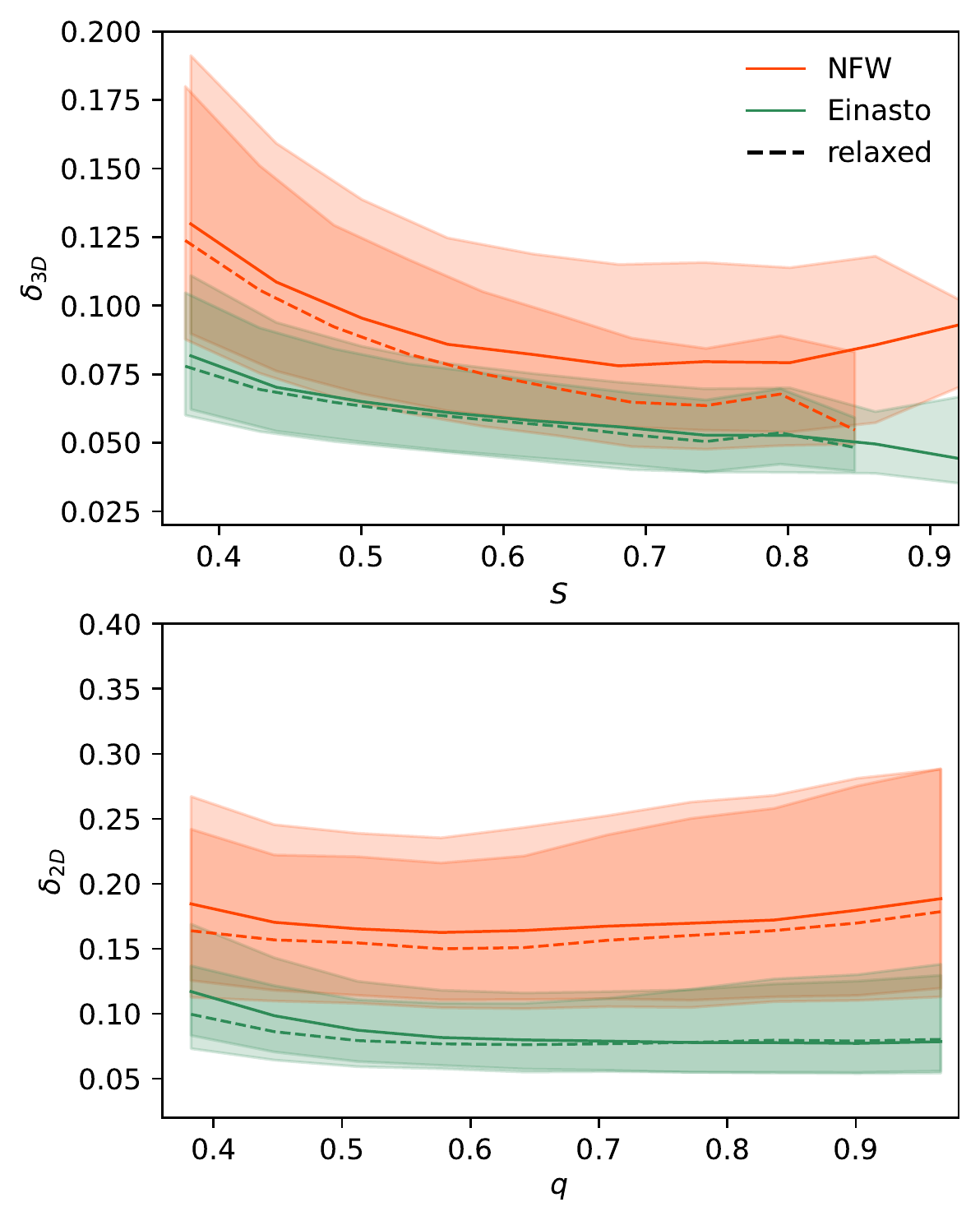}
    \caption{Median 3D and 2D residuals obtained in standard 3D halo sphericities (upper panel) and 2D semi-axis ratios (lower panel) bins, by fitting NFW (orange line) and Einasto profiles (green line), for halos with ellipsoidal densities $100 < \rho^E/\rho_c < 150$. Shadow regions enclose from 25th up to 75th of the residual distributions.}
    \label{fig:res_relax_shape}
\end{figure}

Here we evaluate how well halo radial density distributions are modelled according to the adopted profiles, NFW and Einasto. In order to do that, we select the halos with similar ellipsoidal physical densities. In Fig. \ref{fig:density} we show the comparison between spherical and ellipsoidal 3D halo densities ($\rho$ and $\rho^E$, respectively). Spherical densities are obtained according to the total number of particles enclosed within a spherical volume determined within a radius $r_\text{max}$. On the other hand, ellipsoidal densities are obtained taking into account an ellipsoidal volume determined according to the standard semi-axis ratios. Spherical densities are lower, specially for more prolate halos, since for these halos the total spherical density is underestimated. 

To evaluate the relation between the fitting residuals and the halo properties, we select the halos with $100 < \rho^E/\rho_c < 150$ to neglect the impact of different particle densities in the fitting procedure. By selecting the halos with similar ellipsoidal densities instead of considering the spherical ones, allows to evaluate the impact of the halos shapes in residuals. Moreover it ensures that we are comparing halos with similar radial densities since it properly takes into account the halo shapes. 

In Fig. \ref{fig:res} we present the median values of the residual distributions in bins of the centre offsets, $r_c/r_\text{MAX}$, and the virial energy ratio, $2K/|W|$. In general, residuals are low and in agreement with other studies \citep{Meneghetti2014}, indicating that derived profiles are well described by the fitted models. To evaluate this, we can consider as reference the residuals according to the shot-noise introduced by the mass resolution. Taking this into account, the expected residuals in the density profiles are $\sim 0.1$ and $\sim 0.2$ for 3D and 2D profiles, respectively.
For all derived density distributions, the Einasto profile shows the lowest residual values, offering a better description for the dark matter radial distribution. There is a correlation between the residuals and the centre offset as well as the energy ratio. Therefore and as stated in previous studies, fitting performance highly depends on the halo relaxation state. When selecting relaxed halos, residuals for halos with similar FOF masses and redshifts, show in median similar residuals (Fig- \ref{fig:res_relax}). A subtle trend is observed for higher redshift when adopting an NFW model, where these halos tend to show higher residual values. 

We also study the relation between residuals in the modelling and the halo shapes, according to the standard 3D and 2D sphericities, $S$ and $q$ (Fig. \ref{fig:MFOF_shape}). Higher residuals are obtained for less spherical halos considering the whole sample as well as relaxed halos. Nevertheless, this trend is almost negligible when considering projected profiles. The increase in the dispersion of 2D residuals for rounder halos is related to a poorly constraint of the concentration parameter that is better illustrated in 
\ref{subsec:3D2D}.

\subsection{Relation between FOF and fitted masses}
\label{subsec:fof}
\begin{figure}
    \includegraphics[scale=0.6]{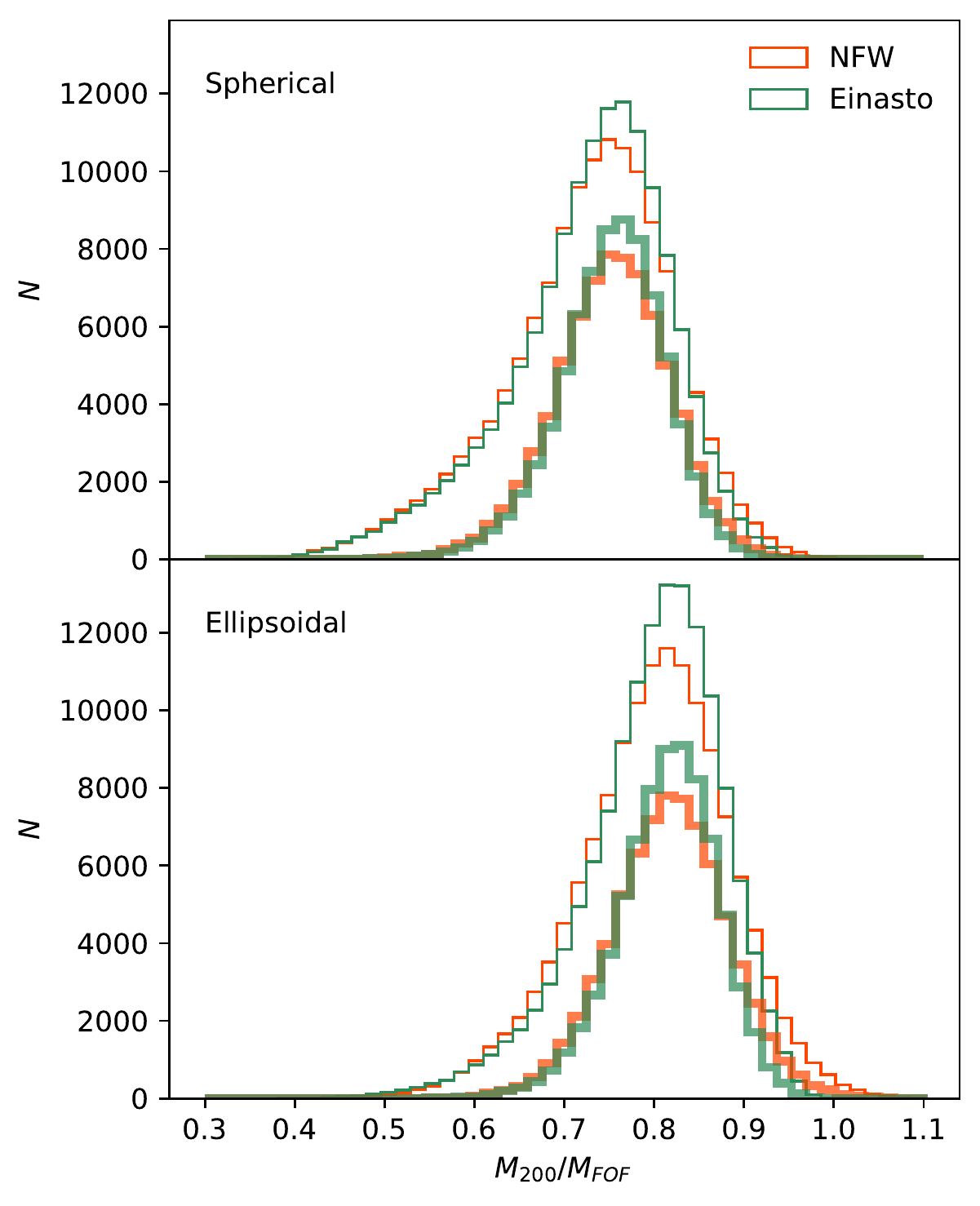}
    \caption{$M_{200}/M_\text{FOF}$ ratio distributions for $M_{200}$ fitted using NFW (red) and Einasto (green) modelling. Thicker  lines correspond to halos classified as relaxed, which show higher ratio values. Upper and lower panels correspond to distributions using $M_{200}$ masses determined using spherical and ellipsoidal radial bins, respectively.}
    \label{fig:MFOF_dist}
\end{figure}


\begin{figure*}
    \includegraphics[scale=0.6]{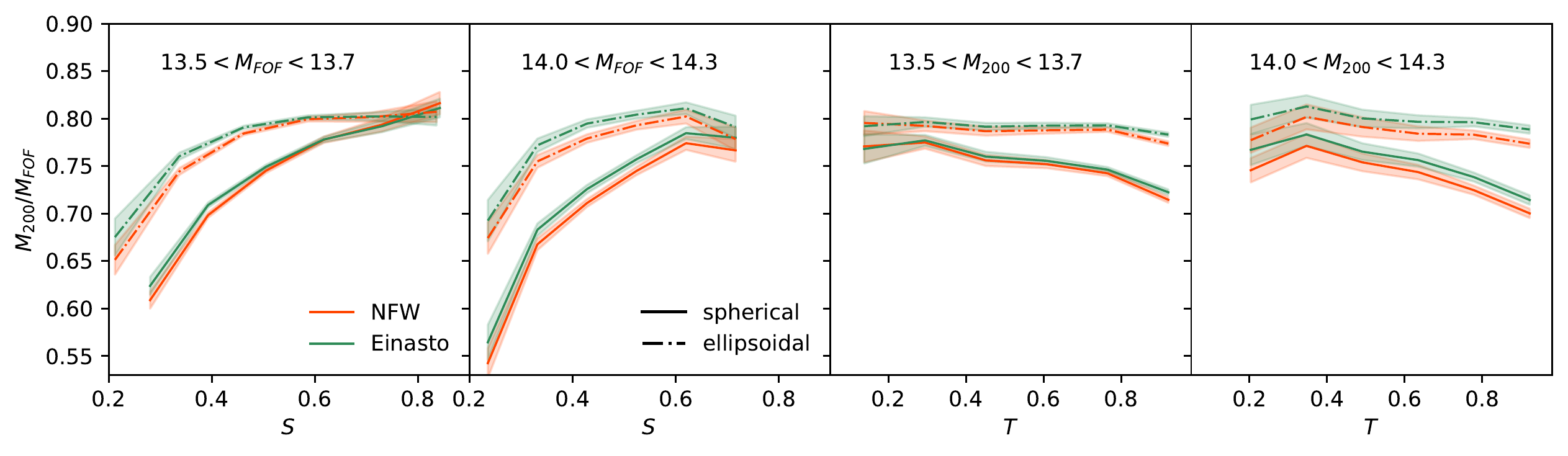}
    \caption{Mean $M_{200}/M_\text{FOF}$ ratios obtained in bins of standard shape parameters sphericity and triaxiality. $M_{200}$ are obtained by fitting density profiles obtained using spherical (solid line) and ellipsoidal (dot-dashed lines) radial bins, using NFW (red) and Einasto (green) models. Relations are obtained for different mass bins including halos within a redshift range $0.35 < z < 0.40$. Shadow regions correspond to RMS errors.}
    \label{fig:MFOF_shape}
\end{figure*}

Differences between $M_\text{FOF}$ and fitted $M_{200}$ are expected to be related with the halo number of particles due to the mass resolution, as well as the profile density concentration, given that for halos with higher concentrations within a fixed mass bin, the isodensity contours will move towards smaller radii, making the FOF mass smaller \citep{Lukic2009}. We show in Fig. \ref{fig:MFOF_dist} the distribution of ratios $M_{200}/M_\text{FOF}$ for the total sample of halos and relaxed ones, taking into account $M_{200}$ masses derived using spherical and ellipsoidal bins. As it can be noticed when considering relaxed halos, the ratio follows a normal distribution and similar distributions are obtained for Einasto as well as for NFW masses. It can be also noticed that when ellipsoidal profiles are taken into account, $M_{200}/M_\text{FOF}$ distributions have higher mean values and slightly lower dispersion. The observed median ratios of $0.75$ using a spherical binning is in agreement with \citep{Lukic2009} values, taking into account that we include all the halos regardless of the density concentrations. 

The observed differences in the $M_{200}/M_\text{FOF}$ ratio distributions when considering spherical and ellipsoidal bins can be better understood if we inspect the relation between FOF and $M_{200}$ masses with respect to the 3D halo shapes. In Fig. \ref{fig:MFOF_shape} we show the median values of the ratio $M_{200}/M_\text{FOF}$ in bins of the halo sphericity and triaxiality. Here we only include relaxed halos in two mass bins and within a redshift range of $0.35 < z < 0.40$ to ensure similar profile resolutions. However, presented results are similar for the whole redshift range. As it can be noticed, when density profiles are computed using spherical bins, less-spherical and prolate halos show lower ratio values. On the other hand, when considering fitted masses using ellipsoidal bins, except for highly elongated halos, the ratios are almost invariant with respect to shape parameters, leading to a mostly constant $M_{200}/M_\text{FOF} \sim 0.8$. Therefore, the differences in the distributions presented in Fig. \ref{fig:MFOF_dist} are the result of a more constrained relation between these masses, which is obtained when halo shapes are considered in modelling the density distributions. 

\subsection{Spherical vs elliptical density distributions}
\label{subsec:ellip}

\begin{figure}
    \includegraphics[scale=0.6]{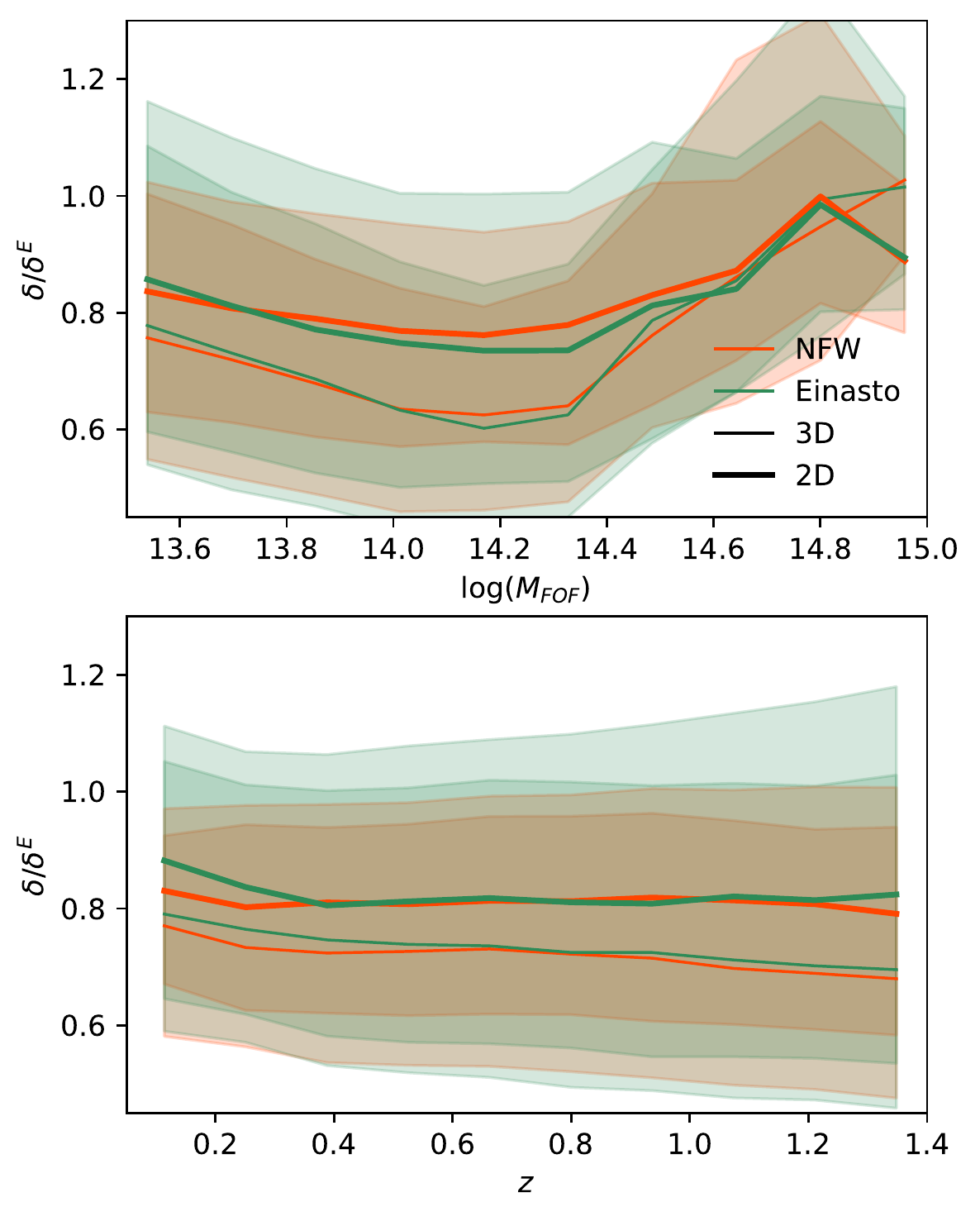}
    \caption{Median residual ratios of fitted 3D and surface density distributions (solid thinner and thicker lines respectively), using spherical and ellipsoidal radial bins, in bins of mass and redshift. We consider halos with ellipsoidal densities $100 < \rho^E/\rho_c < 150$. Shadow regions enclose from 25th up to 75th of the residual distributions.}
    \label{fig:resellip}
\end{figure}

As shown previously, the use of an ellipsoidal binning to compute profiles result in a lower dispersion of constrained masses when comparing them with FOF masses. Here we evaluate how the fitted parameters are affected when an ellipsoidal distribution is considered instead of the usual spherical binning. In Fig. \ref{fig:resellip} we compare the residuals derived from fitting spherical and ellipsoidal profiles only for relaxed halos. As it can be noticed, profiles are better modelled when taking into account a spherical binning, regardless of the halo mass and redshift. This result could be related to the fact that these models are derived using stacked spherical distributions and, although they offer a suitable description when considering ellipsoidal profiles, they better describe the spherical ones.

\begin{figure}
    \includegraphics[scale=0.6]{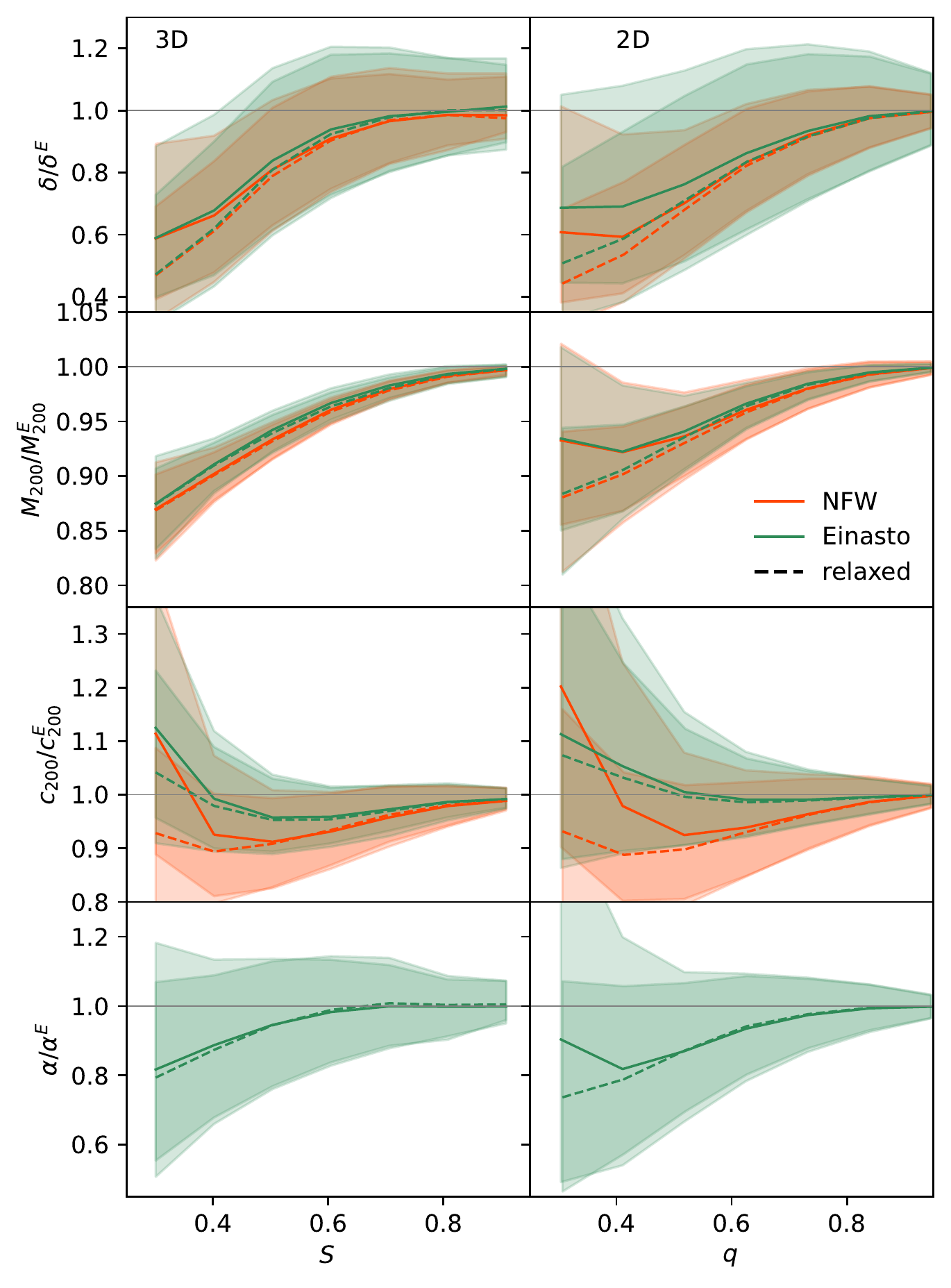}
    \caption{Median residual and fitted parameter ratios of fitted 3D and surface density distributions (left and right panels respectively), using spherical and ellipsoidal radial bins, in terms of the standard elongation, sphericity in 3D and the semi-axis ratio in projection. Shadow regions enclose from 25th up to 75th of the residual distributions.}
    \label{fig:ellip_compare}
\end{figure}

We further inspect the differences in the fitted parameters for ellipsoidal and spherical profiles. In Fig. \ref{fig:ellip_compare} we show the comparison of the fitted parameters in bins of the sphericity and the projected semi-axis ratio. Ellipsoidal residuals tend to be higher than the spherical ones, even when considering more elongated halos an ellipsoidal model does not offer a better description of the density distribution. When using ellipsoidal profiles fitted masses and concentrations tend to be higher. This is expected since we are taking into account the shape of the dark matter particle distribution in computing the profiles, thus, we are constraining the bins to follow the main directions of this distribution. Fitted mass ratios are similar when using NFW and Einasto. On the other hand, fitted concentrations show higher differences for NFW than Einasto, being Einasto fitted concentration almost invariant to the profile binning in projection regardless of the halos shape. However, differences in the profile shape impact on the constrained $\alpha$ parameter, obtaining higher values when ellipsoidal binning are considered. 

\subsection{Radial density concentrations}
\label{subsec:con}
\begin{figure*}
    \includegraphics[scale=0.6]{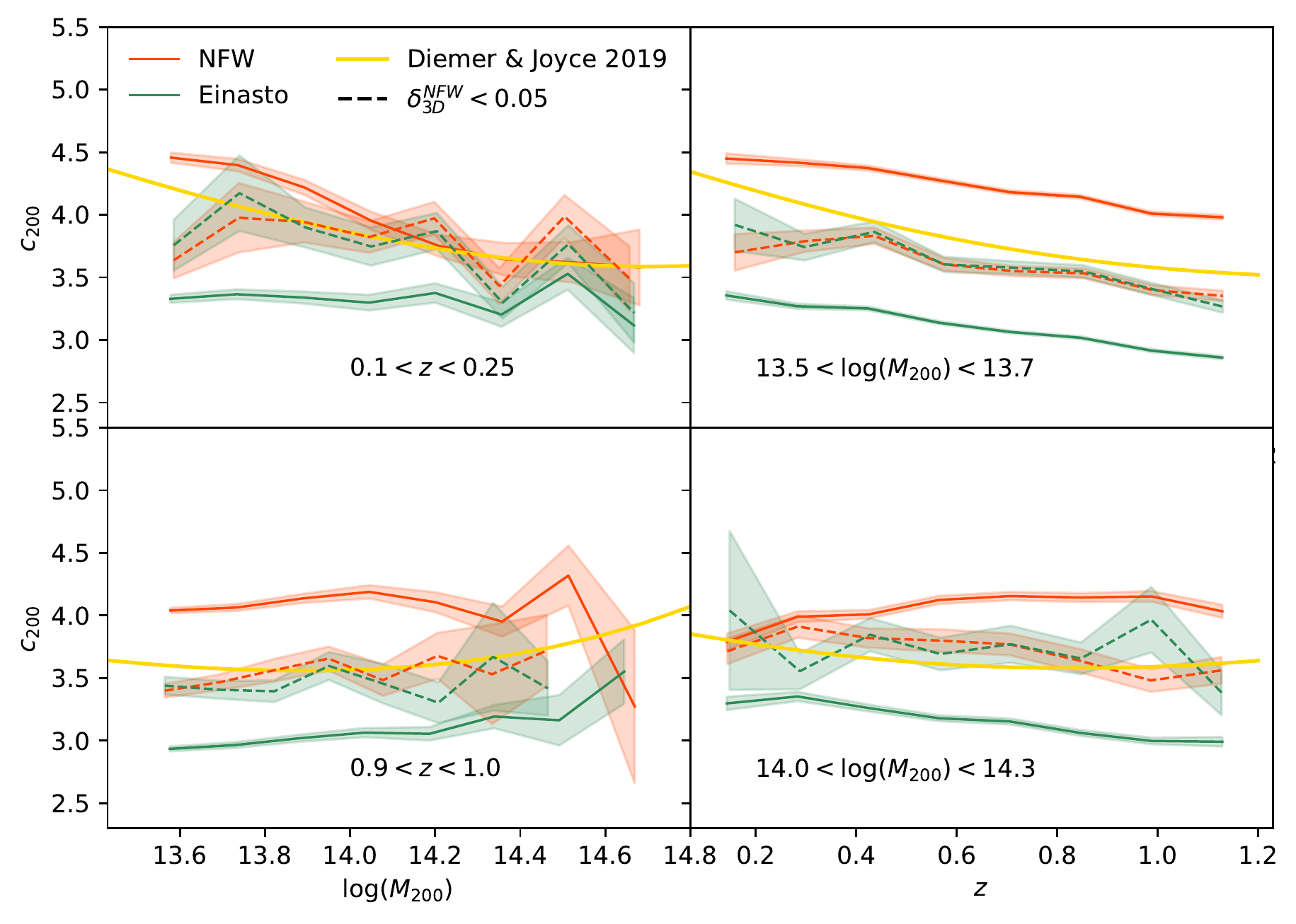}
    \caption{Mean concentration values derived in bins of mass and redshift, by fitting NFW (red lines) and Einasto (green lines) models to the spherical density distributions. Dashed lines correspond to halos with fitting NFW residuals lower than $0.05$. Yellow lines correspond to the relation derived by \citet{Diemer2019}. Shadow regions correspond to RMS errors.}
    \label{fig:c200}
\end{figure*}

\begin{figure*}
    \includegraphics[scale=0.6]{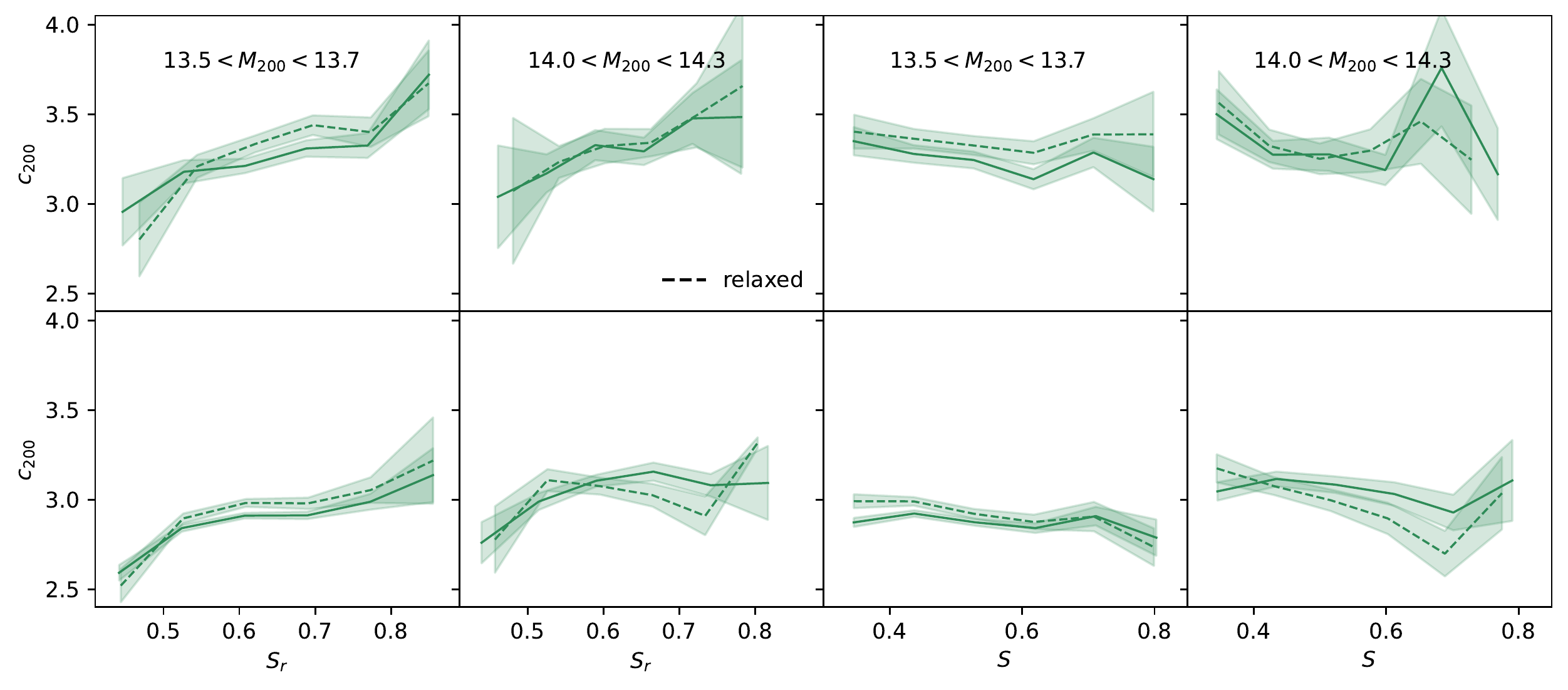}
    \caption{Mean Einasto concentrations in mass bins (columns) for halos within a redshift range of $(0.1,0.25)$ and $(0.9,1.0)$ in the upper and lower row, respectively. Mean values are obtained in terms of the standard and reduced halos sphericity. Shadow regions correspond to RMS errors.}
    \label{fig:c200_shape}
\end{figure*}

There is a well-known relation between the halo mass, redshift and concentration that is linked with the time at which the main halo progenitor assembles a certain fraction of their mass \citep{Zhao2009,Giocoli2012}. Nevertheless, this relation is complex and depends on the adopted cosmology. Moreover, it is not clear how the concentration evolves as a function of time. We inspect the halo mass-redshift-concentration relation by considering the fitted concentrations for relaxed halos in different redshift and mass bins. Results are shown in Fig. \ref{fig:c200}. Fitted concentrations are in general lower for the Einasto profile in agreement with previous studies \citep{Dutton2014,Meneghetti2014}.

We compare the observed trend in mass and redshift with different concentration models available in the literature. Our results show a good correspondence on the observed concentration relations with mass and redshift with the presented by \citep{Diemer2019} but when restricting the sample to those halos which density profiles are well fitted by an NFW model, according to a cut in the residuals ($\delta_\text{NFW} < 0.05$). 

A scatter of the mass and concentration relation is reported, specially when considering non-relaxed halos. This scatter can be related to the different halo formation times \citep{Jing2000,Neto2007,Maccio2008}. Taking this into account, we further inspect the concentration relation but considering the halos shapes in different mass and redshift bins. In order to do this, we use the concentrations estimated using Einasto fitting. Although similar results are obtained using NFW concentrations, as discussed previously, these estimates are more biased with shape due to the radial range in which the fitting procedure is performed. We show in Fig. \ref{fig:c200_shape} the mean concentration in bins of the sphericity parameter. Differences are observed when considering reduced and standard definitions. In particular, there is a clear relation between the concentration and the reduced sphericity for all mass and redshift bins considered, where rounder halos tend to show higher concentration values. This trend is observed for the total sample of halos as well as relaxed ones. The observed trend is steeper when considering halos at lower redshifts. The obtained result is in agreement with the halo formation scenario in which as the dark matter in halos collapse, the density distributions tend to be more concentrated and rounder. It is important to highlight that this is mainly observed when considering reduced shape parameters. Since this sphericity is obtained by up-weighting the inner dark matter particle distribution, it is expected to be also more related to the halo concentration. 

\subsection{Three-dimensional and projected fitted parameters}
\label{subsec:3D2D}
\begin{figure}
    \includegraphics[scale=0.6]{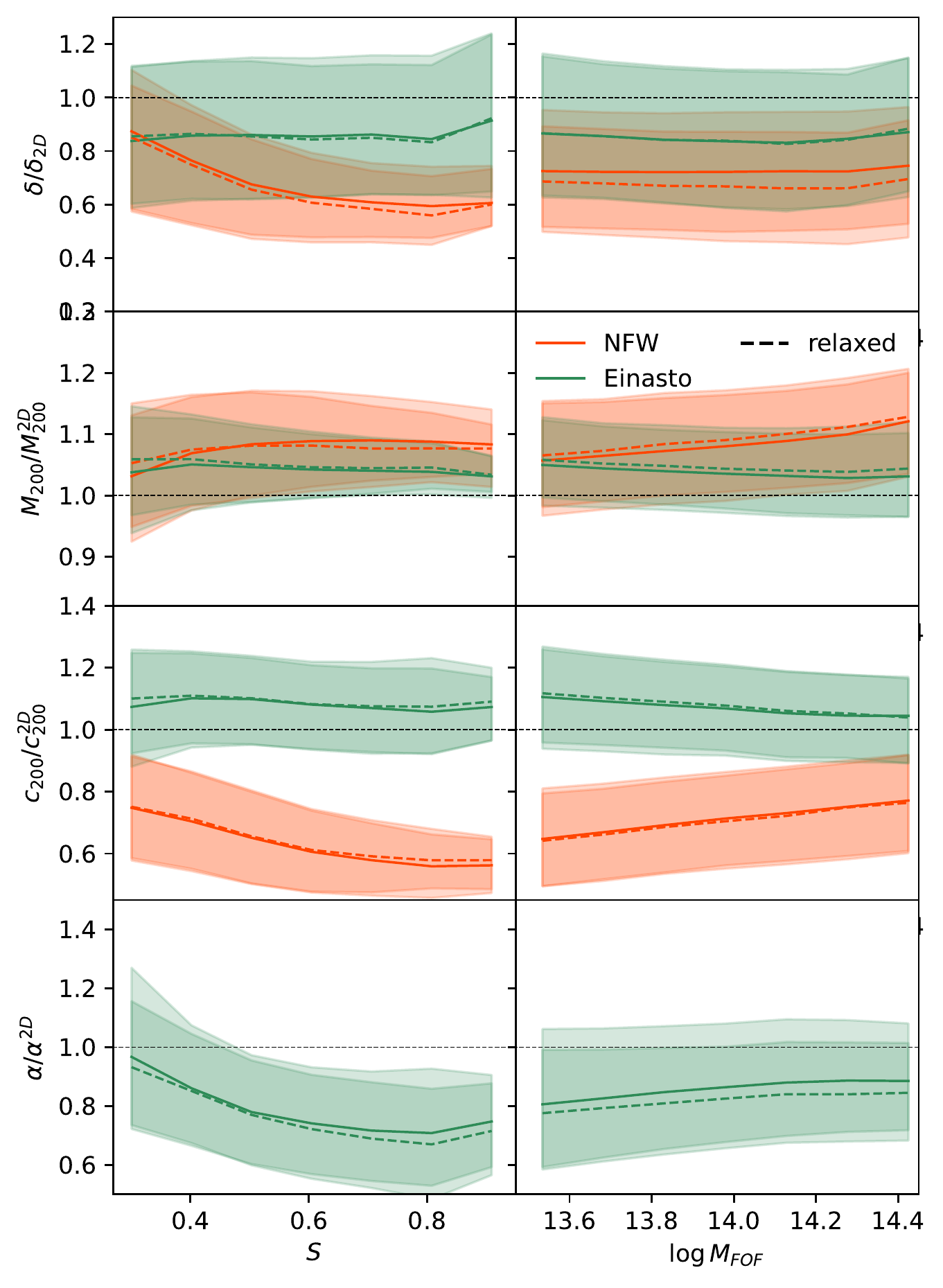}
    \caption{Median residual and fitted parameter ratios between three dimensional and surface density radial distributions, in bins of the standard sphericity and FOF masses. Median ratios are obtained using the total sample of fitted halos (solid line) and the ones classified as relaxed (dashed line). Shadow regions enclose from 25th up to 75th of the residual distributions.}
    \label{fig:3d2d}
\end{figure}

With the scope of analysing potential biases that can be introduced when comparing observed projected quantities with the predicted three-dimensional values obtained from numerical simulations, we analyse the differences between fitted spherical and surface density distributions. Reported differences are often related to the halo shape and the fitted radial range. We show in Fig. \ref{fig:3d2d} the median ratios of fitted parameters in 3D and in projection, in bins of the sphericity and the FOF mass. In general, according to the residuals, fitted profiles have larger residuals when considering the surface density distributions. Masses derived from fitted projected distributions are $\sim 7\%$ and $\sim 4\%$ lower than derived from the three-dimensional density distribution, using NFW and Einasto modelling, respectively. In general, Einasto fitted masses and concentrations biases are constant regardless the halo shape and mass. However, fitted projected $\alpha$ parameters are in median higher for rounder halos. 

\begin{figure*}
    \includegraphics[scale=0.6]{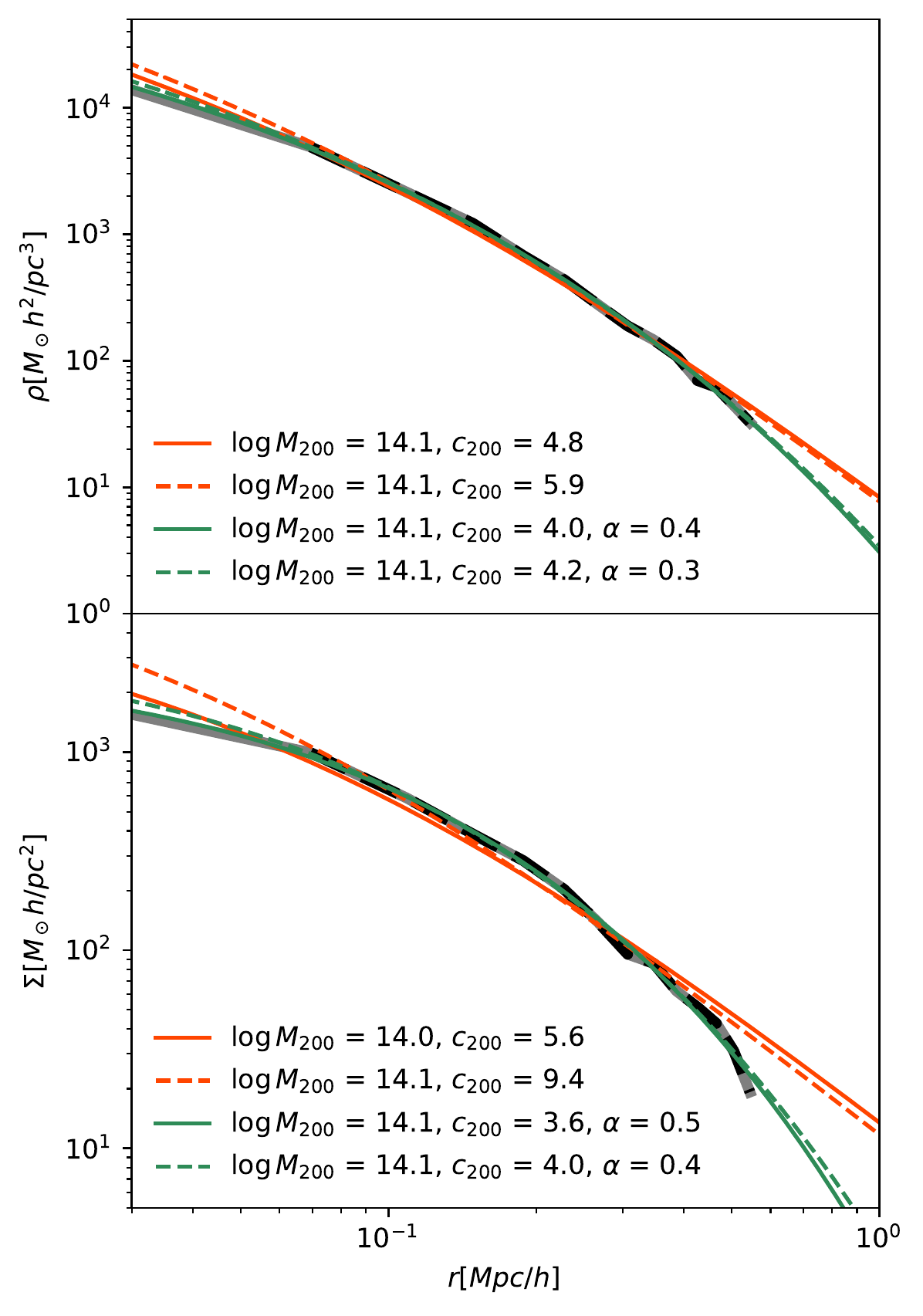}
    \includegraphics[scale=0.6]{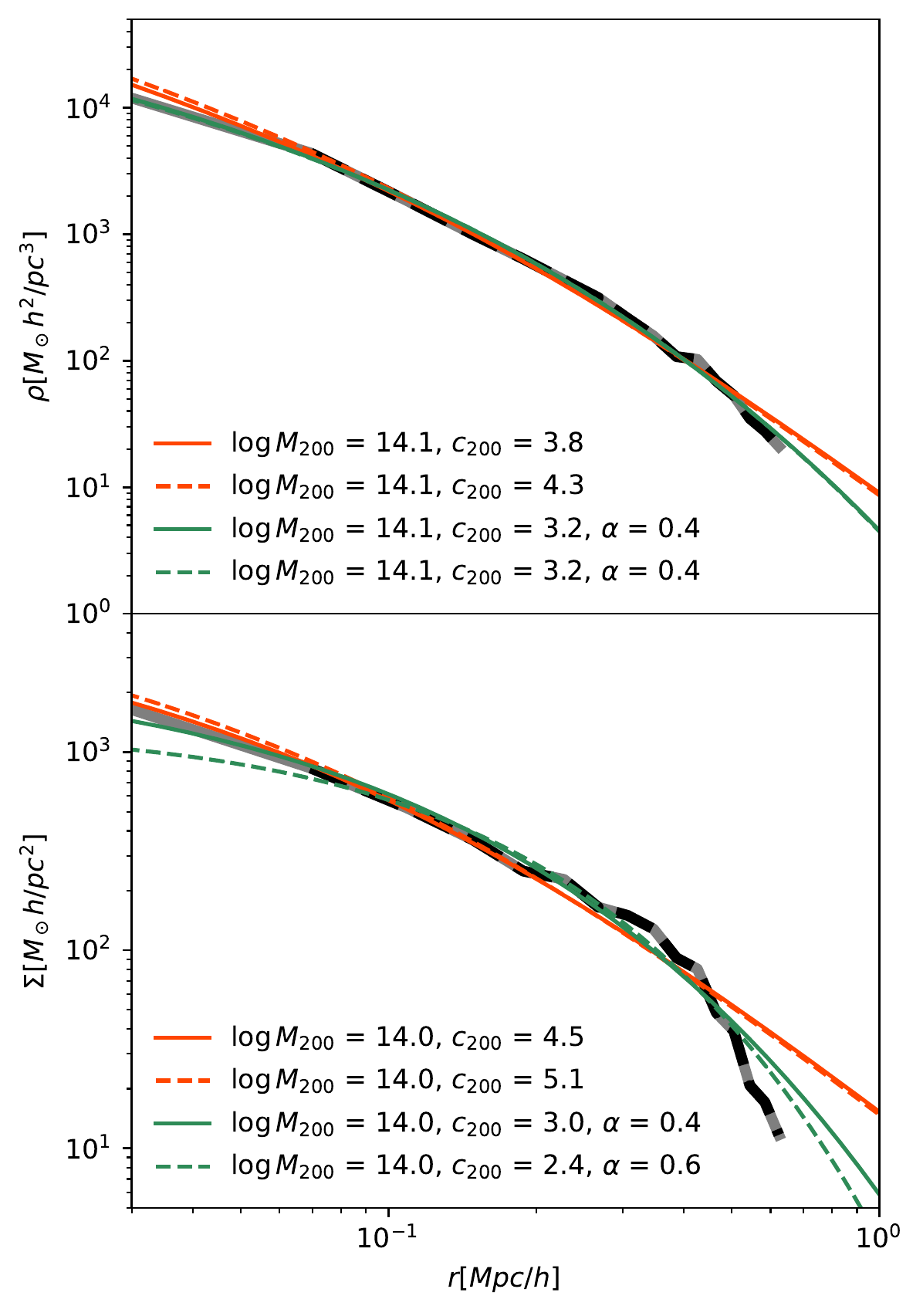}
    \caption{Profile 3D (upper panel) and surface density (lower panel) distributions for two halos with $\log M_\text{FOF} = 14.1$ located at $z = 1.32$ (left panel) and $z = 1.36$ (right panel) and with sphericities of 0.65 and 0.40, respectively. Derived profiles are shown in grey and dashed black lines, within different inner radius cut of  $0.01$ and $0.05 h^{-1}$\,Mpc, respectively. Fitted NFW and Einasto profiles are obtained for the whole radial range (fitted grey profile) showed in solid lines, while also obtained from $0.05 h^{-1}$\,Mpc in dashed lines (fitted dash-black profile). Fitted parameters are shown in each panel.}
    \label{fig:profile}
\end{figure*}

Fitted NFW concentrations are significantly biased to higher values when fitting surface density profiles. These differences increase with halo sphericity, where rounder halos tend to present higher projected NFW concentrations. The observed differences are the result of the interplay between the fitted radial range and the projected shapes, where more elongated halos have a sharpness slope in projection at the outskirts. In Fig. \ref{fig:profile} we show derived profiles for two relaxed halos of similar mass and redshift ($\log M_\text{FOF} = 14.1$ located at $z = 1.32$ and $z = 1.36$), but with different sphericity values ($s = 0.65$ and $s = 0.40$). We show the corresponding fitted NFW and Einasto parameters for two different radial ranges ($r = [0.01,1] h^{-1}$Mpc and $r = [0.05,1] h^{-1}$Mpc). As it can be noticed, the Einasto model provides a more robust description of the radial density distribution concentration regardless of the adopted radial range, as also shown by \citet{Gao2008}. Nevertheless, the differences driven by projection effects are encoded in the third fitted parameter, $\alpha$.

Our results show that the introduced biases when obtaining density parameters from the surface density distribution are mainly related to the fitted radial range. In particular, the impact on the halo shapes in these bias is also related to the fitted range. Higher differences are observed in concentration when an NFW profile is adopted to model the density distribution. Although derived concentrations using Einasto modelling are more robust, $\alpha$ is also overestimated when fitting  2D distributions.

\subsection{Fitted density parameters compared with lensing estimates}

\begin{table} 
 \caption{Selected samples of halos to compute the lensing analysis}
    \centering
    \begin{tabular}{c c c}
    \hline
    \hline
Sample name & $\log M_\text{FOF}$ & $z$ \\
 \\
\hline
HM-Lz & $[14.0,14.5)$  &  $[0.10,0.20)$  \\
LM-Lz & $[13.5,14.0)$  &  $[0.10,0.20)$  \\
HM-Hz & $[14.0,14.5)$  &  $[0.30,0.35)$  \\
LM-Hz & $[13.5,14.0)$  &  $[0.30,0.35)$  \\
\hline
\end{tabular}
\begin{flushleft}
Columns: (1) Sample name of stacked halos; (2) and (3) FOF masses and redshift range of the selected halos in each sample.
\end{flushleft}
    \label{tab:lens_samp}
\end{table}

\begin{table*} 
 \caption{Comparison between lensing parameter estimates and the fitted 3D spherical distributions}
    \centering
    \begin{tabular}{c c c c c c c c c}
    \hline
    \hline
Sample name  &  \multicolumn{3}{c}{NFW} & \multicolumn{4}{c}{Einasto} \\
            & $\tilde{M}_{200}/\langle M_{200} \rangle$ & $\tilde{c}_{200}/\langle c_{200} \rangle$ & $\delta$ 
            & $\tilde{M}_{200}/\langle M_{200} \rangle$ & $\tilde{c}_{200}/\langle c_{200} \rangle$  
            & $\tilde{\alpha} / \langle \alpha \rangle$ & $\delta$ &  $\tilde{c}_{200}/\langle c^{Ein}_{200} \rangle$ \\
\hline
\multicolumn{9}{c}{all halos} \\
\hline
HM-Lz  & $1.05$ & $0.81$ & $0.013$ & $1.06$ & $0.84$ & $0.88$ & $0.006$ & $0.93$ \\
LM-Lz  & $1.02$ & $0.81$ & $0.023$ & $1.04$ & $0.81$ & $0.91$ & $0.010$ & $1.03$ \\
HM-Hz  & $1.00$ & $0.76$ & $0.026$ & $1.02$ & $0.76$ & $1.10$ & $0.015$ & $0.92$ \\
LM-Hz  & $1.01$ & $0.75$ & $0.033$ & $1.02$ & $0.75$ & $0.98$ & $0.021$ & $0.96$ \\
\hline
\multicolumn{9}{c}{only relaxed} \\
\hline
HM-Lz  & $1.04$ & $0.85$ & $0.021$ & $1.05$ & $0.82$ & $1.06$ & $0.010$ & $0.97$ \\ 
LM-Lz  & $1.03$ & $0.83$ & $0.027$ & $1.05$ & $0.82$ & $0.94$ & $0.012$ & $1.05$ \\ 
HM-Hz  & $1.01$ & $0.77$ & $0.024$ & $1.03$ & $0.76$ & $1.08$ & $0.016$ & $0.93$ \\ 
LM-Hz  & $1.00$ & $0.82$ & $0.035$ & $1.03$ & $0.75$ & $1.08$ & $0.022$ & $1.05$ \\ 
\hline
\multicolumn{9}{c}{only relaxed - elongated ($q < 0.6$)} \\
\hline
HM-Lz & $0.97$ & $0.76$ & $0.029$ & $1.00$ & $0.72$ & $1.24$ & $0.013$ & $0.87$ \\ 
LM-Lz & $0.99$ & $0.78$ & $0.040$ & $1.01$ & $0.77$ & $1.05$ & $0.017$ & $1.05$ \\ 
HM-Hz & $0.96$ & $0.74$ & $0.029$ & $0.98$ & $0.73$ & $1.11$ & $0.019$ & $0.91$ \\ 
LM-Hz & $0.95$ & $0.73$ & $0.045$ & $0.98$ & $0.72$ & $1.07$ & $0.024$ & $1.00$ \\ 
\hline
\multicolumn{9}{c}{only relaxed - rounder ($q \geq 0.6$)} \\
\hline
HM-Lz & $1.08$ & $0.91$ & $0.018$ & $1.08$ & $0.89$ & $0.96$ & $0.010$ & $1.03$ \\ 
LM-Lz & $1.06$ & $0.81$ & $0.026$ & $1.07$ & $0.86$ & $0.84$ & $0.014$ & $0.99$ \\ 
HM-Hz & $1.05$ & $0.80$ & $0.026$ & $1.08$ & $0.77$ & $1.08$ & $0.020$ & $0.95$ \\ 
LM-Hz & $1.03$ & $0.83$ & $0.035$ & $1.05$ & $0.78$ & $1.03$ & $0.025$ & $1.01$ \\ 
\hline
\multicolumn{9}{c}{only relaxed - rounder and triaxial ($q \geq 0.6$, $T < 0.8$)} \\
\hline
HM-Lz & $1.04$ & $0.83$ & $0.016$ & $1.04$ & $0.89$ & $0.85$ & $0.011$ & $0.96$ \\ 
LM-Lz & $1.04$ & $0.78$ & $0.031$ & $1.05$ & $0.80$ & $0.91$ & $0.015$ & $0.94$ \\ 
HM-Hz & $1.03$ & $0.79$ & $0.028$ & $1.05$ & $0.76$ & $1.17$ & $0.021$ & $0.92$ \\ 
LM-Hz & $0.99$ & $0.84$ & $0.032$ & $1.01$ & $0.79$ & $0.99$ & $0.024$ & $1.01$ \\ 
\hline
\multicolumn{9}{c}{only relaxed - rounder and prolate ($q \geq 0.6$, $T \geq 0.8$)} \\
\hline
HM-Lz & $1.14$ & $1.08$ & $0.026$ & $1.18$ & $0.88$ & $1.24$ & $0.016$ & $1.20$ \\ 
LM-Lz & $1.10$ & $0.89$ & $0.020$ & $1.10$ & $1.03$ & $0.70$ & $0.017$ & $1.13$ \\ 
HM-Hz & $1.10$ & $0.82$ & $0.026$ & $1.12$ & $0.82$ & $0.98$ & $0.022$ & $1.01$ \\ 
LM-Hz & $1.09$ & $0.87$ & $0.038$ & $1.12$ & $0.79$ & $1.10$ & $0.027$ & $1.08$ \\ 
\hline
\end{tabular}
\begin{flushleft}
Columns: (1) Sample name of stacked halos; (2), (3) $M_{200}$ mass 
and $c_{200}$ concentrations fitted from the density contrast profile adopting an NFW model, scaled according to the mean of the
fitted 3D spherical density distributions; (4) residuals of the contrast density profile adopting an NFW model; (5), (6) and (7) $M_{200}$ mass, $c_{200}$ concentration and $\alpha$ fitted from the density contrast profile adopting an Einasto model, scaled according to the mean of the fitted 3D spherical density distributions; (8) residuals of the contrast density profile adopting an Einasto model; 
(9) lensing concentration fitted according a NFW profile scaled by the mean of the distribution of fitted Einasto concentrations from 3D density distributions.
\end{flushleft}
    \label{tab:lens_res}
\end{table*}

\begin{figure*}
    \includegraphics[scale=0.6]{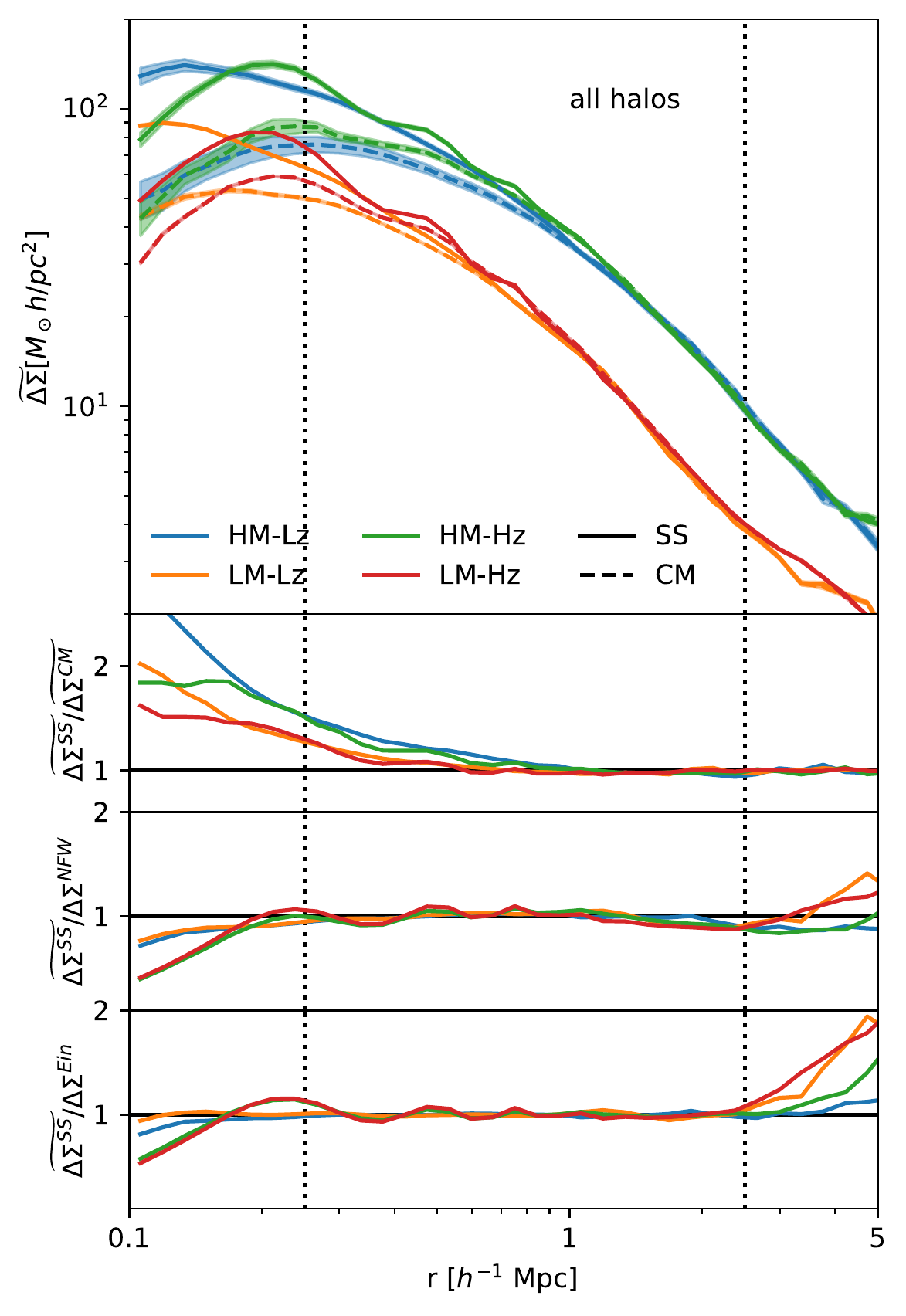}
    \includegraphics[scale=0.6]{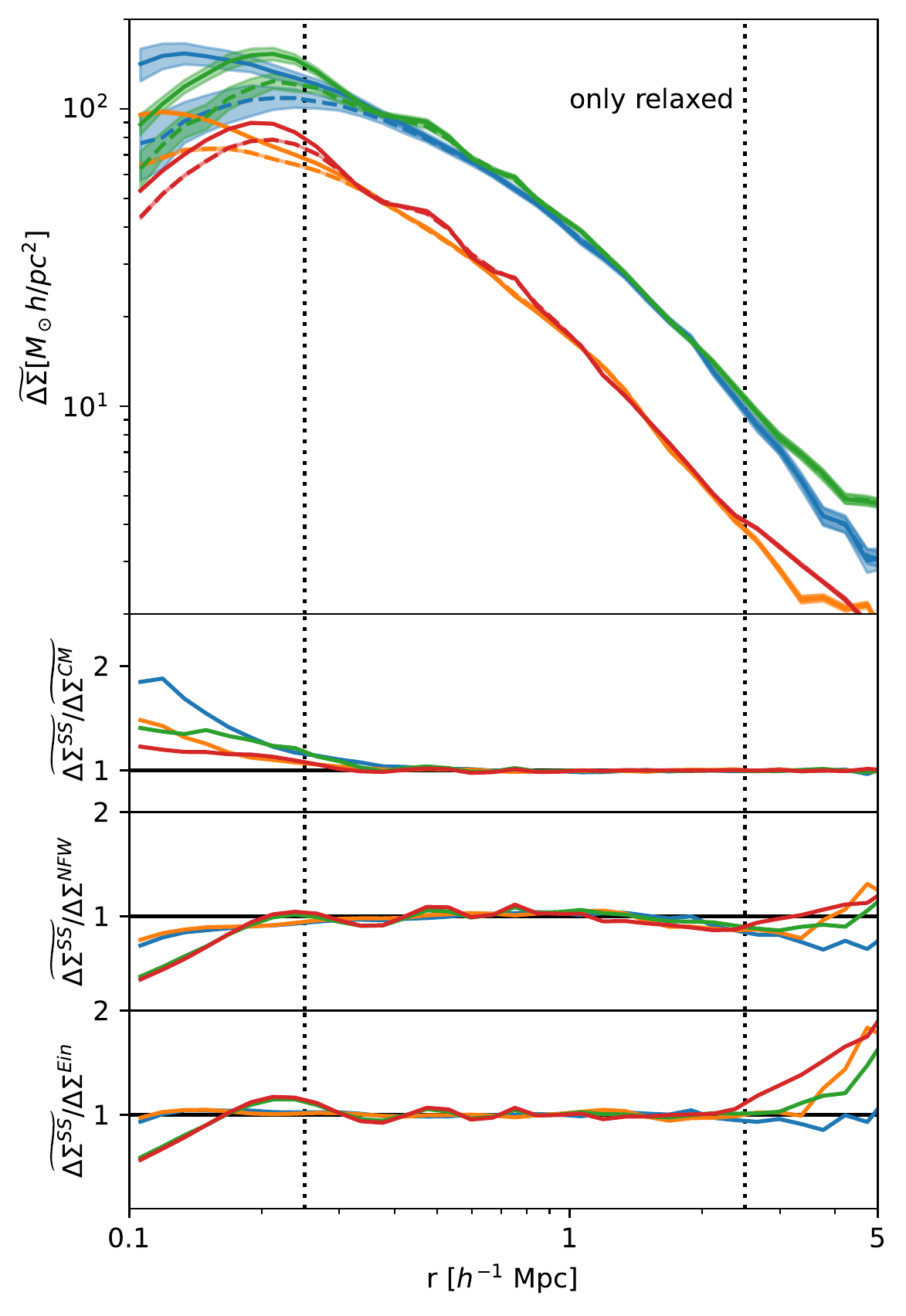}
    \caption{Left and right panels show the density contrast for all the halos and only relaxed ones, respectively. The colour code corresponds to the four samples of halos defined in Table \ref{tab:lens_samp} selected according to the FOF mass and redshift. Upper panels show the computed density contrast profiles ($\widetilde{\Delta \Sigma}$) computed according to the \textit{shear} estimator (Eq. \ref{eq:DSigma}), taking into account the halo shrinking-sphere ($SSC$, solid lines) and the FOF centre of mass ($CM$, dashed lines). The shadow regions correspond to the adopted errors computed using a jacknife resampling. The second row of panels correspond to the profile ratio using the two centre definitions ($\widetilde{\Delta \Sigma^{SSC}} / \widetilde{\Delta \Sigma^{CM}}$). Third and forth row show the ratio between the computed profile adopting the shringking sphere centre and the best fit NFW and Einasto model, $\widetilde{\Delta \Sigma^{SSC}} / \Delta \Sigma^{NFW}$ and $\widetilde{\Delta \Sigma^{SSC}} / \Delta \Sigma^{Ein}$, respectively.}
    \label{fig:profile_lens}
\end{figure*}

\begin{figure*}
    \includegraphics[scale=0.6]{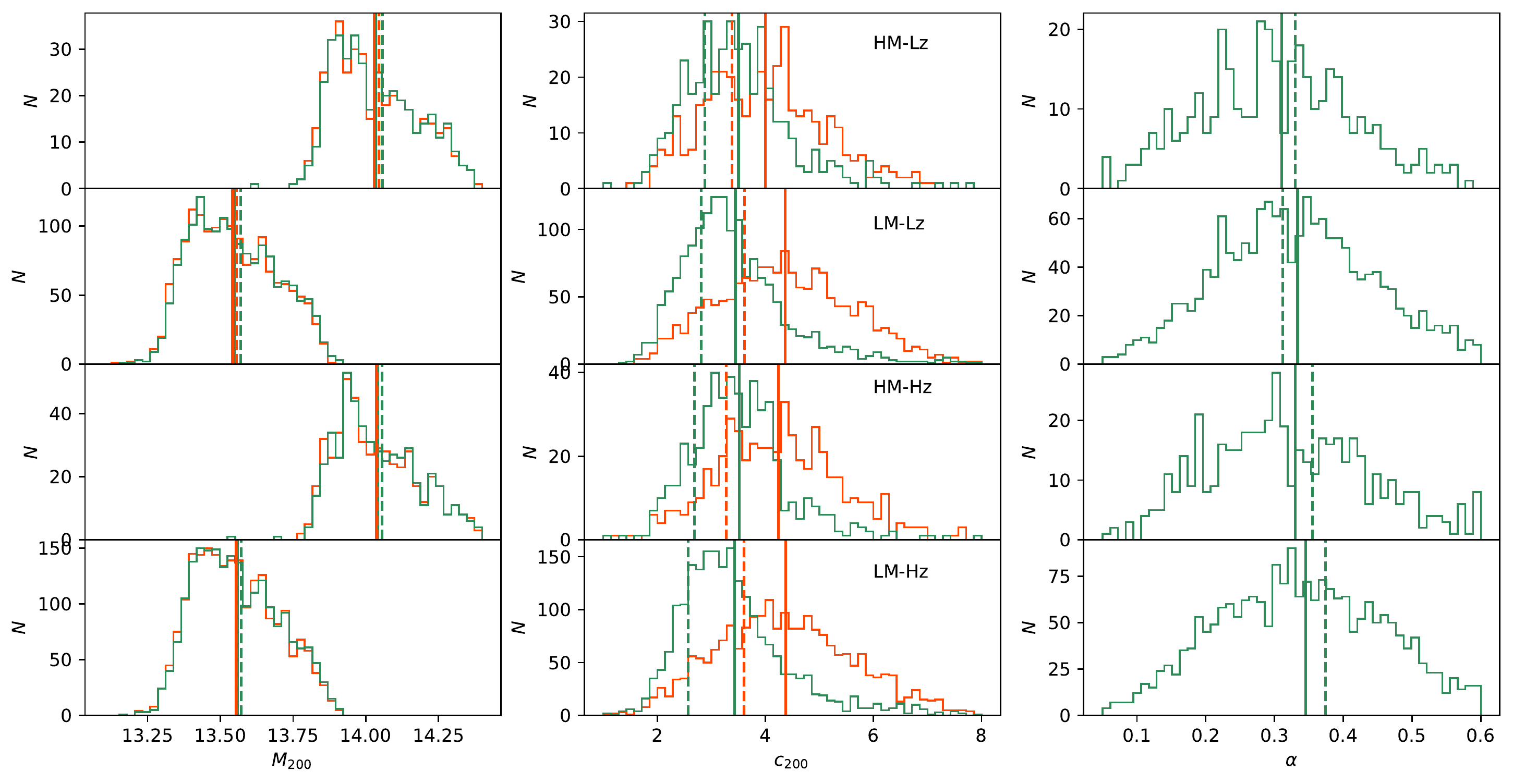}
    \caption{NFW (orange) and Einasto (green) parameter distributions fitted from the 3D radial density profiles for the selected halo samples defined in Table \ref{tab:lens_samp}. Solid lines indicate the mean value of each distribution while dashed lines correspond to the fitted values obtained using the stacked-lensing techniques. }
    \label{fig:dist_lens}
\end{figure*}

Taking advantage of the lensing properties provided for MICE simulation through the mock catalogue, we can test the performance of the fitted profiles using weak-lensing stacking techniques. The estimation of halo properties using these techniques, in particular halo shape determinations, will be deeply inspected in a posterior paper (Gonzalez et al. in prep.). Here we compare lensing masses and concentrations estimates with the mean values of the distributions of fitted parameters for four samples of stacked halos, selected according to $M_\text{FOF}$ and $z$ (see Table \ref{tab:lens_samp}). We summarise how the lensing estimates are derived while the detailed analysis can be found in Gonzalez et al. in prep. A similar study that computes the halo masses using weak-lensing stacking techniques was also presented in \citet{Rodriguez2020}.

To obtain lensing masses and concentrations we compute the density contrast profiles as:
\begin{equation} \label{eq:DSigma}
 \widetilde{\Delta \Sigma}(r) = \frac{\sum_{j=1}^{N_L} \sum_{i=1}^{N_{S,j}} \Sigma_{{\rm crit},ij} \gamma_{{\rm t},ij}}{\sum_{j=1}^{N_L} N_{S,j} },
\end{equation}
where $\gamma_{{\rm t},ij}$ is the tangential \textit{shear} component for the $i-$th source of the $j-$th lens, which is related with the shape distortion caused by the deflection of the light due to the gravitational lensing effect. $N_L$ is the number of halos considered as lenses for the stacking and $N_{S,j}$ the number of sources located at a projected physical distance $r \pm \delta r$ from the adopted $j-$th halo centre. $\Sigma_{{\rm crit},ij}$ is the critical density defined as:
\begin{equation} \label{eq:sig_crit}
\Sigma_{\rm{crit}} = \dfrac{c^{2}}{4 \pi G} \dfrac{D_{i}}{D_{j} D_{ij}},
\end{equation}
where $D_{i}$, $D_{j}$ and $D_{ij}$ are the angular diameter distances to the lens, to the source and from the lens to the source, respectively, computed according to the source and lenses redshifts ($z_{S}$ and $z_L$). The sample of halos considered as lenses for the stacking are those that satisfy the mass and redshift cuts detailed in Table \ref{tab:lens_samp}, while the source galaxies are selected as those that fulfil $z_{S} > z_{L} + 0.1$. Profiles are computed considering 40 non-overlapping concentric logarithmic annuli, from $100\,h^{-1}$\,kpc up to $10.0\,h^{-1}$\,Mpc.

In Fig. \ref{fig:profile_lens} we show the profiles for the halo samples, computed considering as the halo centre the shrinking-sphere as well as the FOF centre of mass. When adopting the FOF centre of mass, the profile depletes in the inner regions given that this centre definition does not necessary correspond with the peak of the density distribution. This effect is more significant for higher mass and non-relaxed systems. It can be also noticed a depletion in the inner region even when adopting the shrinking-sphere centre. The reduction of the lensing signal in the inner regions is related with the noise introduced by numerical effects, since the shear is associated with all the integrated mass distribution within the considered radius \citep{Oguri2011}. Also, the inner depletion is introduced due to the angular resolution of the lensing parameters, thus affecting more to halos selected at higher redshifts. For MICE, the angular resolution of the \textit{shear} is $0.43$ arcmin, which corresponds to $\sim 50\,h^{-1}$\,kpc and $\sim 90\,h^{-1}$\,kpc for the lower and higher redshift stacked halo samples respectively.

We obtain the parameters that describe the density distributions by fitting the $\Delta \Sigma$ profiles computed adopting the shrinking-sphere centre. We fit NFW and Einasto models using Colossus with the same methodology as for fitting the density distributions (see \ref{subsec:fitdens}) within a radial range from $250\,h^{-1}$\,kpc up to $2500\,h^{-1}$\,kpc and using a jacknife resampling to compute the errors. The upper limit is selected to avoid the contribution of the neighbouring mass distribution, known as the 2-halo term, while the inner limit in the radial range is selected to avoid the numerical effects introduced in the profile and considering the simulation resolution. The election of neglecting the neighbouring mass contribution is based on simplicity. However, the inclusion of the 2-halo term is properly described in Gonzalez et al. (in prep.) and we highlight that the observed bias in the fitted parameters are not highly affected by this choice. Fitting residuals are defined in analogous to Eq. \ref{eq:res3d} and \ref{eq:res2d} as:
\begin{equation}
    \delta = \frac{1}{N_\text{doff}} \sum_i [\log \widetilde{\Delta \Sigma_i} - \log \Delta \Sigma(r_i,\textbf{p})]^2,
\end{equation}
where $\widetilde{\Delta \Sigma}$ is the computed density contrast profile defined in Eq. \ref{eq:DSigma} and $\Delta \Sigma$ is the best fitted NFW and Einasto density contrasts parametrized through \textbf{p}. 

In Table \ref{tab:lens_res} we show fitted lensing estimates scaled by the mean values of the parameter distributions obtained according to the 3D dark-matter particle density profiles and the residuals. In order to avoid the effect of outliers, we discard the halos with fitted concentrations and $\alpha$ parameters higher than $10$ and $0.7$, respectively. NFW and Einasto parameters obtained from the 3D density distributions together with the fitted lensing estimates for relaxed halos are shown in Fig. \ref{fig:dist_lens}. There are no significant differences when considering non-relaxed halos in the analysis. This reinforce the centre choice together with the fact that the stacking lensing procedure results in a softening of the surface density distribution, thus, reducing the effect introduced by the presence of substructure. Fitted NFW and Einasto scaled values are in agreement and, tanking into account the low values of $\delta$, it follows that both models adequately describe the computed density contrast profiles. This is also evidenced by the ratio between the lensing profiles and the fitted models shown in Fig. \ref{fig:profile_lens}, which are mainly in agreement within the fitted range. However, and consistently with the trends analysed, residuals are in general lower when adopting Einasto modelling. 

Estimated lensing masses and $\alpha$ parameters are all in agreement with the average fitted values from the 3D density distribution within $\lessapprox 5 \%$ and $\lessapprox 10 \%$, respectively. On the other hand, a significant bias is obtained for concentrations, being lensing estimates $\sim 20 \%$ lower than mean 3D fitted concentrations. This bias is consistent with the trend reported in \ref{subsec:3D2D} for Einasto concentrations. Bias of fitted parameters have been already extensively reported in many previous studies based in numerical simulations \citep[e.g.][]{Oguri2011,Bahe2012,Meneghetti2014,Sereno2016,Child2018} and are related with the methodology adopted to measure concentration as the adopted fitting range, modelling, halo selection algorithm, individual fitting vs. stacked analysis, among other possible things as  projection effects introduced by the halo triaxility. We evaluate the impact of halo triaxiality in the observed biases by repeating the analysis for the rounder ($q \geq 0.6$)  and elongated ($q < 0.6$) halos, for these sub-samples the introduced biases affects in opposite directions, obtaining lower masses and concentrations for more elongated halos. We further split the sub-sample of rounder halos according to the triaxiality parameter, to discard those that are highly elongated along the line-of-sight by considering $T < 0.8$. For this sub-sample the observed trends are similar than when considering the total sample of relaxed halos, thus the observed biases might mainly related with the applied fitting methodology than with projections effects introduced by the halos deviations from spherical symmetry. Additionally, we show the results for highly elongated along the line-of-sight halos ($q \geq 0.6$, $T \geq 0.8$). This sub-sample show the highest deviations for lensing masses from the ones measured using the 3D density profiles. 

We notice that NFW lensing concentrations are in agreement within a $\lessapprox 5 \%$ with Einasto concentration distributions for individual halos (last column of Table \ref{tab:lens_res}). This result is not expected to have a physical motivation but to be related with a bias compensation that result in similar concentrations. We point out this result just as useful check when using the presented catalogue for lensing comparisons.  

\color{black}

\section{Summary and conclusions}
\label{sec:summary}

The study of the halo properties, such as masses, shapes and concentrations, in 3D as well as in 2D projection is necessary to evaluate the potential biases in cosmological tests that are introduced when comparing observations to predictions from theory or simulations.
In particular, halo shapes introduce important systematic effects since they are not spherical (most are prolate).
In this paper, we presented a halo catalogue 
containing the 
main properties derived in this work for dark-matter halos identified in MICE-GC simulation, which is accessible through the CosmoHub platform\footnote{\href{https://cosmohub.pic.es/home}{https://cosmohub.pic.es/home}} \citep{Carretero2017,TALLADA2020100391}. The properties include two centre determinations based on the centre of mass and the shrinking-sphere algorithm, gravitational self-potential and kinetic energies and density profiles (see Appendix \ref{app:catalog} for the column description). Density profiles were computed using spherical as well as ellipsoidal bins in order to consider the halo shapes in modelling the radial distribution and fitted using NFW and Einasto profiles. We also analysed the density distributions of a subsample of cluster-size halos. We study the interplay between the halo shapes, relaxation state and fitted parameters from the radial density distributions.

We inspect the mass ratio, $M_{200}/M_\text{FOF}$, considering the fitted $M_{200}$ masses. Ratio distributions are similar for NFW as well as for Einasto modelling, with median values of $\sim 0.75$ when considering spherical radial binning to compute the profiles. We observe a dependence between the mass ratio and halo shapes, being lower for less-spherical and more prolate halos. When ellipsoidal profiles are taken into account, the median of $M_{200}/M_\text{FOF}$ distribution is higher ($0.81$) and this ratio shows a lower dependence on halo shapes. The observed result indicates that FOF masses are related with the ellipsoidal overdensity mass, generalising the result presented by \citet{Lukic2009} in which the relation is studied assuming spherical symmetry. 

Although the profiles computed using a ellipsoidal radial binning are obtained considering the morphology of the particle distribution, the adopted models, according to the residuals, do not offer a better description for these distributions. This result can be related to the fact that NFW and Einasto models are derived assuming spherical distributions. Also, it is important to take into account that ellipsoidal profiles are obtained assuming a constant semi-axis ratio across the whole radial range. As shown in previous studies \citep[e.g.][]{Schneider2012,Despali2017}, dark matter particle distribution shapes vary with the radial range and standard parameters can be affected by external substructures.  

According to the fitted concentrations we inspect their relation with mass, redshift and halo shapes. We found a good agreement between the concentration, mass, redshift relations with the relation presented by \citet{Diemer2019}, but only when restricting the halo sample to those well described by an NFW model. We obtain a clear relation between reduced sphericity and halo concentrations, where less-spherical halos have lower concentrations. This result is in agreement with the halo formation scenario in which, as the matter in halos collapses and times goes by, they become rounder and more concentrated \citep[e.g.][]{Allgood2006,Despali2017,Munoz2011,Okoli2017}. We conclude that the observed dispersion for a particular mass bin in halo concentrations could be related to the halo formation time through their shapes. 

We also study the relation between fitted 3D and 2D density parameters. Since we use the same dark-matter particles of the halos to obtain the surface density distribution, we can neglect the effect of line-of-sight substructure as expected in observational studies and that could have a significant impact on observable estimates. Thus, our study allows to evaluate the impact of modelling in recovering density parameters alone. We obtain a mass bias of $7\%$ and $5\%$ for NFW and Einasto modelling regardless of the halo shapes. This supports the recent study presented in \citet{Debackere2022} who propose using projected aperture masses for cluster cosmology since 3D cluster masses
cannot be directly compared without assuming a density distribution profile. This assumption introduces model-dependent biases and uncertainties due to the mismatch between the adopted and the true cluster density profile. Fitted concentrations are also biased when they are determined using surface instead of 3D density distributions, but in different ways when adopting NFW and Einasto modelling. The observed differences between 3D and 2D derived parameters are mainly related to the adopted radial range to perform the fitting procedure. The inner limit of the radial range in this work is fixed at $50h^{-1}$\,kpc taking into account the softening length. This imposes a more restrictive lower limit in terms of the virial radius for lower mass halos. Thus, for these halos concentrations are poorly constrained  given the lack of information at the inner regimes. This is mainly important when adopting NFW modelling given that derived concentrations for this model are more susceptible to the choice of the radial range. On the other hand, fitted Einasto concentrations in 3D and in projection show lower differences, but still an almost constant bias related to mass and sphericity. 

Finally we compare fitted parameters with lensing estimates derived by fitting density contrast profiles for stacked samples of halos. Presented shrinking sphere centres significantly improve the lensing profile at the inner regions. Although the mean masses of stacked halos are adequately recovered using the lensing analysis, determined concentrations are in general biased. The obtained bias is comparable with the observed when comparing 3D with 2D fitted parameters. We further inspect these biases by considering different sub-samples selected according to the halo shape parameters. According to this study, we conclude that the observed biases in concentrations are mainly related with the differences in the fitting procedure rather than projection effects due to the halo triaxiality. Nevertheless, lensing masses can be indeed highly overestimated using the lensing approach for line-of-sight elongated halos, which is in agreement with previous studies \citep[e.g.,][]{Gavazzi2005,Limousin2013,Osato2018,Zhang2022}.  

The presented catalogue aims to provide useful parameters that describe in detail the density distributions and relaxation state of the halos available in MICE-GC simulation. This enables several analysis that can be conducted in order to establish observational constraints for the mass distribution resulted from the adopted cosmological paradigm. In particular, the catalogue provides a centre definition consistent with the peak density distribution that allow a shape estimate of the dark-matter particle configuration less affected by the external substructure.
This is also useful for several studies as lensing analysis, that are highly susceptible to the centre definition. Furthermore, we provide a well constrained mass that can be linked with fitted observable for cluster-size halos, together with other fitted properties that characterise the dark-matter density distributions. 

\section*{Acknowledgements}

\section*{Data Availability}

Computed parameters are available through CosmoHub platform\footnote{\href{https://cosmohub.pic.es/home}{https://cosmohub.pic.es/home}} \citep{Carretero2017,TALLADA2020100391} under the name `MICE HALO properties'. Computations are based on the dark matter particles associated with the identified halos which are available under request evaluated by the MICE collaboration team.



\bibliographystyle{mnras}
\bibliography{example} 




\appendix
\label{app:catalog}
\section{Detailed columns of the halo properties catalogue}

In this section we specify the columns of the released catalogue. Additionally to the properties presented in \ref{sec:prop}, we compute and  the specific angular momentum vectors that can be related with the halo spin and which is defined as:
\begin{equation}
    \vec{j} = \frac{\sum_i^{N_p} \vec{r}_i \times \vec{v}_i}{N_p},
\end{equation}
where $\vec{r}_i$ and $\vec{v}_i$ are the positional and velocity vectors of the $i-$th halo dark-matter particle with respect to the FOF centre of mass. The sums runs over the whole number of identified particles linked in the FOF halo. This parameter is provided in the catalogue and left for future analysis 

\onecolumn

\small
\begin{longtable}[c]{l l l l}
\caption{Description of the halo properties catalogue}\\
 \hline
Column name & Format & Description & Units \\
\hline
halo\_id & int & halo id number & \\
Npart & bigint & number of FOF particles & \\
lgM & float & log base 10 of FoF halo mass & \\
xc & float & comoving FOF centre of mass distance (x-component) & (Mpc/h) \\
yc & float & comoving FOF centre of mass distance (y-component) & (Mpc/h) \\
zc & float & comoving FOF centre of mass distance (z-component) & (Mpc/h) \\
xc\_rc & float & comoving FOF shrink-sphere centre distance (x-component) & (Mpc/h) \\
yc\_rc & float & comoving FOF shrink-sphere centre distance (y-component) & (Mpc/h) \\
zc\_rc & float & comoving FOF shrink-sphere centre distance (z-component) & (Mpc/h) \\
redshift & float & true redshift & \\
r\_max & float & maximum radial distance from the FOF centre of mass to the most distant FOF particle & \\
vxc & float & physical (peculiar) velocity of FOF centre of mass  (x-component) & (km/s) \\
vyc & float & physical (peculiar) velocity of FOF centre of mass (y-component) & (km/s) \\
vzc & float & physical (peculiar) velocity of FOF centre of mass (z-component) & (km/s) \\
Jx & float & angular momentum vector (x-component) & (Mpc/h * km/s) \\
Jy & float & angular momentum vector (y-component) & (Mpc/h * km/s) \\
Jz & float & angular momentum vector (z-component) & (Mpc/h * km/s) \\
Ekin & float & Kinetic energy & (10$^{30}$ kg*m$^2$/s$^2$) \\
Epot & float & Gravitational energy & (10$^{30}$ kg*m$^2$/s$^2$) \\
q2D\_CM & float & 2D standard semi-axis ratio with respect to the CM centre & \\
a2Dx\_CM & float & 2D standard eigenvector associated to the major semi-axis with respect to the CM centre (ra-component) & \\
a2Dy\_CM & float & 2D standard eigenvector associated to the major semi-axis with respect to the CM centre (dec-component) & \\
b2Dx\_CM & float & 2D standard eigenvector associated to the minor semi-axis with respect to the CM centre (ra-component) & \\
b2Dy\_CM & float & 2D standard eigenvector associated to the minor semi-axis with respect to the CM centre (dec-component) & \\
q2Dr\_CM & float & 2D reduced semi-axis ratio with respect to the CM centre & \\
a2Drx\_CM & float & 2D reduced eigenvector associated to the major semi-axis with respect to the CM centre (ra-component) & \\
a2Dry\_CM & float & 2D reduced eigenvector associated to the major semi-axis with respect to the CM centre (dec-component) & \\
b2Drx\_CM & float & 2D reduced eigenvector associated to the minor semi-axis with respect to the CM centre (ra-component) & \\
b2Dry\_CM & float & 2D reduced eigenvector associated to the minor semi-axis with respect to the CM centre (dec-component) & \\
q3D\_CM & float & 3D standard medium to major semi-axis ratio with respect to the CM centre & \\
s3D\_CM & float & 3D standard minor to major semi-axis ratio with respect to the CM centre & \\
a3Dx\_CM & float & 3D standard eigenvector associated to the major semi-axis with respect to the CM centre (x-component) & \\
a3Dy\_CM & float & 3D standard eigenvector associated to the major semi-axis with respect to the CM centre (y-component) & \\
a3Dz\_CM & float & 3D standard eigenvector associated to the major semi-axis with respect to the CM centre (z-component) & \\
c3Dx\_CM & float & 3D standard eigenvector associated to the minor semi-axis with respect to the CM centre (x-component) & \\
c3Dy\_CM & float & 3D standard eigenvector associated to the minor semi-axis with respect to the CM centre (y-component) & \\
c3Dz\_CM & float & 3D standard eigenvector associated to the minor semi-axis with respect to the CM centre (z-component) & \\
q3Dr\_CM & float & 3D reduced medium to major semi-axis ratio with respect to the CM centre & \\
s3Dr\_CM & float & 3D reduced minor to major semi-axis ratio with respect to the CM centre & \\
a3Drx\_CM & float & 3D reduced eigenvector associated to the major semi-axis with respect to the CM centre (x-component) & \\
a3Dry\_CM & float & 3D reduced eigenvector associated to the major semi-axis with respect to the CM centre (y-component) & \\
a3Drz\_CM & float & 3D reduced eigenvector associated to the major semi-axis with respect to the CM centre (z-component) & \\
c3Drx\_CM & float & 3D reduced eigenvector associated to the minor semi-axis with respect to the CM centre (x-component) & \\
c3Dry\_CM & float & 3D reduced eigenvector associated to the minor semi-axis with respect to the CM centre (y-component) & \\
c3Drz\_CM & float & 3D reduced eigenvector associated to the minor semi-axis with respect to the CM centre (z-component) & \\
a2D\_SSC & float & 2D standard eigenvalue associated to the major semi-axis with respect to the SSC centre & (kpc/h) \\
b2D\_SSC & float & 2D standard eigenvalue associated to the minor semi-axis with respect to the SSC centre & (kpc/h) \\
a2Dx\_SSC & float & 2D standard eigenvector associated to the major semi-axis with respect to the SSC centre (ra-component) & \\
a2Dy\_SSC & float & 2D standard eigenvector associated to the major semi-axis with respect to the SSC centre (dec-component) & \\
b2Dx\_SSC & float & 2D standard eigenvector associated to the minor semi-axis with respect to the SSC centre (ra-component) & \\
b2Dy\_SSC & float & 2D standard eigenvector associated to the minor semi-axis with respect to the SSC centre (dec-component) & \\
a2Dr\_SSC & float & 2D reduced eigenvalue associated to the major semi-axis with respect to the SSC centre & (kpc/h) \\
b2Dr\_SSC & float & 2D reduced eigenvalue associated to the minor semi-axis with respect to the SSC centre & (kpc/h) \\
a2Drx\_SSC & float & 2D reduced eigenvector associated to the major semi-axis with respect to the SSC centre (ra-component) & \\
a2Dry\_SSC & float & 2D reduced eigenvector associated to the major semi-axis with respect to the SSC centre (dec-component) & \\
b2Drx\_SSC & float & 2D reduced eigenvector associated to the minor semi-axis with respect to the SSC centre (ra-component) & \\
b2Dry\_SSC & float & 2D reduced eigenvector associated to the minor semi-axis with respect to the SSC centre (dec-component) & \\
a3D\_SSC & float & 3D standard eigenvalue associated to the major semi-axis with respect to the SSC centre & (kpc/h) \\
b3D\_SSC & float & 3D standard eigenvalue associated to the medium semi-axis with respect to the SSC centre & (kpc/h) \\
c3D\_SSC & float & 3D standard eigenvalue associated to the minor semi-axis with respect to the SSC centre & (kpc/h) \\
a3Dx\_SSC & float & 3D standard eigenvector associated to the major semi-axis with respect to the SSC centre (x-component) & \\
a3Dy\_SSC & float & 3D standard eigenvector associated to the major semi-axis with respect to the SSC centre (y-component) & \\
a3Dz\_SSC & float & 3D standard eigenvector associated to the major semi-axis with respect to the SSC centre (z-component) & \\
b3Dx\_SSC & float & 3D standard eigenvector associated to the medium semi-axis with respect to the SSC centre (x-component) & \\
b3Dy\_SSC & float & 3D standard eigenvector associated to the medium semi-axis with respect to the SSC centre (y-component) & \\
b3Dz\_SSC & float & 3D standard eigenvector associated to the medium semi-axis with respect to the SSC centre (z-component) & \\
c3Dx\_SSC & float & 3D standard eigenvector associated to the minor semi-axis with respect to the SSC centre (x-component) & \\
c3Dy\_SSC & float & 3D standard eigenvector associated to the minor semi-axis with respect to the SSC centre (y-component) & \\
c3Dz\_SSC & float & 3D standard eigenvector associated to the minor semi-axis with respect to the SSC centre (z-component) & \\
a3Dr\_SSC & float & 3D reduced eigenvalue associated to the major semi-axis with respect to the SSC centre & (kpc/h) \\
b3Dr\_SSC & float & 3D reduced eigenvalue associated to the medium semi-axis with respect to the SSC centre & (kpc/h) \\
c3Dr\_SSC & float & 3D reduced eigenvalue associated to the minor semi-axis with respect to the SSC centre & (kpc/h) \\
a3Drx\_SSC & float & 3D reduced eigenvector associated to the major semi-axis with respect to the SSC centre (x-component) & \\
a3Dry\_SSC & float & 3D reduced eigenvector associated to the major semi-axis with respect to the SSC centre (y-component) & \\
a3Drz\_SSC & float & 3D reduced eigenvector associated to the major semi-axis with respect to the SSC centre (z-component) & \\
b3Drx\_SSC & float & 3D reduced eigenvector associated to the medium semi-axis with respect to the SSC centre (x-component) & \\
b3Dry\_SSC & float & 3D reduced eigenvector associated to the medium semi-axis with respect to the SSC centre (y-component) & \\
b3Drz\_SSC & float & 3D reduced eigenvector associated to the medium semi-axis with respect to the SSC centre (z-component) & \\
c3Drx\_SSC & float & 3D reduced eigenvector associated to the minor semi-axis with respect to the SSC centre (x-component) & \\
c3Dry\_SSC & float & 3D reduced eigenvector associated to the minor semi-axis with respect to the SSC centre (y-component) & \\
c3Drz\_SSC & float & 3D reduced eigenvector associated to the minor semi-axis with respect to the SSC centre (z-component) & \\
r\_0 & float & Mean radial distance to the 1st bin of the density profile & (kpc/h) \\
r\_1 & float & Mean radial distance to the 2nd bin of the density profile & (kpc/h) \\
r\_2 & float & Mean radial distance to the 3rd bin of the density profile & (kpc/h) \\
r\_3 & float & Mean radial distance to the 4th bin of the density profile & (kpc/h) \\
r\_4 & float & Mean radial distance to the 5th bin of the density profile & (kpc/h) \\
r\_5 & float & Mean radial distance to the 6th bin of the density profile & (kpc/h) \\
r\_6 & float & Mean radial distance to the 7th bin of the density profile & (kpc/h) \\
r\_7 & float & Mean radial distance to the 8th bin of the density profile & (kpc/h) \\
r\_8 & float & Mean radial distance to the 9th bin of the density profile & (kpc/h) \\
r\_9 & float & Mean radial distance to the 10th bin of the density profile & (kpc/h) \\
r\_10 & float & Mean radial distance to the 11th bin of the density profile & (kpc/h) \\
r\_11 & float & Mean radial distance to the 12th bin of the density profile & (kpc/h) \\
r\_12 & float & Mean radial distance to the 13th bin of the density profile & (kpc/h) \\
r\_13 & float & Mean radial distance to the 14th bin of the density profile & (kpc/h) \\
r\_14 & float & Mean radial distance to the 15th bin of the density profile & (kpc/h) \\
r\_15 & float & Mean radial distance to the 16th bin of the density profile & (kpc/h) \\
r\_16 & float & Mean radial distance to the 17th bin of the density profile & (kpc/h) \\
r\_17 & float & Mean radial distance to the 18th bin of the density profile & (kpc/h) \\
r\_18 & float & Mean radial distance to the 19th bin of the density profile & (kpc/h) \\
r\_19 & float & Mean radial distance to the 20th bin of the density profile & (kpc/h) \\
r\_20 & float & Mean radial distance to the 21th bin of the density profile & (kpc/h) \\
r\_21 & float & Mean radial distance to the 22th bin of the density profile & (kpc/h) \\
r\_22 & float & Mean radial distance to the 23th bin of the density profile & (kpc/h) \\
r\_23 & float & Mean radial distance to the 24th bin of the density profile & (kpc/h) \\
r\_24 & float & Mean radial distance to the 25th bin of the density profile & (kpc/h) \\
rho\_0 & float & 3D density in spherical shells of the 1st radial bin & ($M_\odot h^2/Mpc^3$) \\
rho\_1 & float & 3D density in spherical shells of the 2nd radial bin & ($M_\odot h^2/Mpc^3$) \\
rho\_2 & float & 3D density in spherical shells of the 3rd radial bin & ($M_\odot h^2/Mpc^3$) \\
rho\_3 & float & 3D density in spherical shells of the 4th radial bin & ($M_\odot h^2/Mpc^3$) \\
rho\_4 & float & 3D density in spherical shells of the 5th radial bin & ($M_\odot h^2/Mpc^3$) \\
rho\_5 & float & 3D density in spherical shells of the 6th radial bin & ($M_\odot h^2/Mpc^3$) \\
rho\_6 & float & 3D density in spherical shells of the 7th radial bin & ($M_\odot h^2/Mpc^3$) \\
rho\_7 & float & 3D density in spherical shells of the 8th radial bin & ($M_\odot h^2/Mpc^3$) \\
rho\_8 & float & 3D density in spherical shells of the 9th radial bin & ($M_\odot h^2/Mpc^3$) \\
rho\_9 & float & 3D density in spherical shells of the 10th radial bin & ($M_\odot h^2/Mpc^3$) \\
rho\_10 & float & 3D density in spherical shells of the 11th radial bin & ($M_\odot h^2/Mpc^3$) \\
rho\_11 & float & 3D density in spherical shells of the 12th radial bin & ($M_\odot h^2/Mpc^3$) \\
rho\_12 & float & 3D density in spherical shells of the 13th radial bin & ($M_\odot h^2/Mpc^3$) \\
rho\_13 & float & 3D density in spherical shells of the 14th radial bin & ($M_\odot h^2/Mpc^3$) \\
rho\_14 & float & 3D density in spherical shells of the 15th radial bin & ($M_\odot h^2/Mpc^3$) \\
rho\_15 & float & 3D density in spherical shells of the 16th radial bin & ($M_\odot h^2/Mpc^3$) \\
rho\_16 & float & 3D density in spherical shells of the 17th radial bin & ($M_\odot h^2/Mpc^3$) \\
rho\_17 & float & 3D density in spherical shells of the 18th radial bin & ($M_\odot h^2/Mpc^3$) \\
rho\_18 & float & 3D density in spherical shells of the 19th radial bin & ($M_\odot h^2/Mpc^3$) \\
rho\_19 & float & 3D density in spherical shells of the 20th radial bin & ($M_\odot h^2/Mpc^3$) \\
rho\_20 & float & 3D density in spherical shells of the 21th radial bin & ($M_\odot h^2/Mpc^3$) \\
rho\_21 & float & 3D density in spherical shells of the 22th radial bin & ($M_\odot h^2/Mpc^3$) \\
rho\_22 & float & 3D density in spherical shells of the 23th radial bin & ($M_\odot h^2/Mpc^3$) \\
rho\_23 & float & 3D density in spherical shells of the 24th radial bin & ($M_\odot h^2/Mpc^3$) \\
rho\_24 & float & 3D density in spherical shells of the 25th radial bin & ($M_\odot h^2/Mpc^3$) \\
rhoE\_0 & float & 3D density in ellipsoidal shells of the 1st radial bin & ($M_\odot h^2/Mpc^3$) \\
rhoE\_1 & float & 3D density in ellipsoidal shells of the 2nd radial bin & ($M_\odot h^2/Mpc^3$) \\
rhoE\_2 & float & 3D density in ellipsoidal shells of the 3rd radial bin & ($M_\odot h^2/Mpc^3$) \\
rhoE\_3 & float & 3D density in ellipsoidal shells of the 4th radial bin & ($M_\odot h^2/Mpc^3$) \\
rhoE\_4 & float & 3D density in ellipsoidal shells of the 5th radial bin & ($M_\odot h^2/Mpc^3$) \\
rhoE\_5 & float & 3D density in ellipsoidal shells of the 6th radial bin & ($M_\odot h^2/Mpc^3$) \\
rhoE\_6 & float & 3D density in ellipsoidal shells of the 7th radial bin & ($M_\odot h^2/Mpc^3$) \\
rhoE\_7 & float & 3D density in ellipsoidal shells of the 8th radial bin & ($M_\odot h^2/Mpc^3$) \\
rhoE\_8 & float & 3D density in ellipsoidal shells of the 9th radial bin & ($M_\odot h^2/Mpc^3$) \\
rhoE\_9 & float & 3D density in ellipsoidal shells of the 10th radial bin & ($M_\odot h^2/Mpc^3$) \\
rhoE\_10 & float & 3D density in ellipsoidal shells of the 11th radial bin & ($M_\odot h^2/Mpc^3$) \\
rhoE\_11 & float & 3D density in ellipsoidal shells of the 12th radial bin & ($M_\odot h^2/Mpc^3$) \\
rhoE\_12 & float & 3D density in ellipsoidal shells of the 13th radial bin & ($M_\odot h^2/Mpc^3$) \\
rhoE\_13 & float & 3D density in ellipsoidal shells of the 14th radial bin & ($M_\odot h^2/Mpc^3$) \\
rhoE\_14 & float & 3D density in ellipsoidal shells of the 15th radial bin & ($M_\odot h^2/Mpc^3$) \\
rhoE\_15 & float & 3D density in ellipsoidal shells of the 16th radial bin & ($M_\odot h^2/Mpc^3$) \\
rhoE\_16 & float & 3D density in ellipsoidal shells of the 17th radial bin & ($M_\odot h^2/Mpc^3$) \\
rhoE\_17 & float & 3D density in ellipsoidal shells of the 18th radial bin & ($M_\odot h^2/Mpc^3$) \\
rhoE\_18 & float & 3D density in ellipsoidal shells of the 19th radial bin & ($M_\odot h^2/Mpc^3$) \\
rhoE\_19 & float & 3D density in ellipsoidal shells of the 20th radial bin & ($M_\odot h^2/Mpc^3$) \\
rhoE\_20 & float & 3D density in ellipsoidal shells of the 21th radial bin & ($M_\odot h^2/Mpc^3$) \\
rhoE\_21 & float & 3D density in ellipsoidal shells of the 22th radial bin & ($M_\odot h^2/Mpc^3$) \\
rhoE\_22 & float & 3D density in ellipsoidal shells of the 23th radial bin & ($M_\odot h^2/Mpc^3$) \\
rhoE\_23 & float & 3D density in ellipsoidal shells of the 24th radial bin & ($M_\odot h^2/Mpc^3$) \\
rhoE\_24 & float & 3D density in ellipsoidal shells of the 25th radial bin & ($M_\odot h^2/Mpc^3$) \\
Sigma\_0 & float & 2D density in circular rings of the 1st radial bin & ($M_\odot h^2/Mpc^2$) \\
Sigma\_1 & float & 2D density in circular rings of the 2nd radial bin & ($M_\odot h^2/Mpc^2$) \\
Sigma\_2 & float & 2D density in circular rings of the 3rd radial bin & ($M_\odot h^2/Mpc^2$) \\
Sigma\_3 & float & 2D density in circular rings of the 4th radial bin & ($M_\odot h^2/Mpc^2$) \\
Sigma\_4 & float & 2D density in circular rings of the 5th radial bin & ($M_\odot h^2/Mpc^2$) \\
Sigma\_5 & float & 2D density in circular rings of the 6th radial bin & ($M_\odot h^2/Mpc^2$) \\
Sigma\_6 & float & 2D density in circular rings of the 7th radial bin & ($M_\odot h^2/Mpc^2$) \\
Sigma\_7 & float & 2D density in circular rings of the 8th radial bin & ($M_\odot h^2/Mpc^2$) \\
Sigma\_8 & float & 2D density in circular rings of the 9th radial bin & ($M_\odot h^2/Mpc^2$) \\
Sigma\_9 & float & 2D density in circular rings of the 10th radial bin & ($M_\odot h^2/Mpc^2$) \\
Sigma\_10 & float & 2D density in circular rings of the 11th radial bin & ($M_\odot h^2/Mpc^2$) \\
Sigma\_11 & float & 2D density in circular rings of the 12th radial bin & ($M_\odot h^2/Mpc^2$) \\
Sigma\_12 & float & 2D density in circular rings of the 13th radial bin & ($M_\odot h^2/Mpc^2$) \\
Sigma\_13 & float & 2D density in circular rings of the 14th radial bin & ($M_\odot h^2/Mpc^2$) \\
Sigma\_14 & float & 2D density in circular rings of the 15th radial bin & ($M_\odot h^2/Mpc^2$) \\
Sigma\_15 & float & 2D density in circular rings of the 16th radial bin & ($M_\odot h^2/Mpc^2$) \\
Sigma\_16 & float & 2D density in circular rings of the 17th radial bin & ($M_\odot h^2/Mpc^2$) \\
Sigma\_17 & float & 2D density in circular rings of the 18th radial bin & ($M_\odot h^2/Mpc^2$) \\
Sigma\_18 & float & 2D density in circular rings of the 19th radial bin & ($M_\odot h^2/Mpc^2$) \\
Sigma\_19 & float & 2D density in circular rings of the 20th radial bin & ($M_\odot h^2/Mpc^2$) \\
Sigma\_20 & float & 2D density in circular rings of the 21th radial bin & ($M_\odot h^2/Mpc^2$) \\
Sigma\_21 & float & 2D density in circular rings of the 22th radial bin & ($M_\odot h^2/Mpc^2$) \\
Sigma\_22 & float & 2D density in circular rings of the 23th radial bin & ($M_\odot h^2/Mpc^2$) \\
Sigma\_23 & float & 2D density in circular rings of the 24th radial bin & ($M_\odot h^2/Mpc^2$) \\
Sigma\_24 & float & 2D density in circular rings of the 25th radial bin & ($M_\odot h^2/Mpc^2$) \\
SigmaE\_0 & float & 2D density in elliptical rings of the 1st radial bin & ($M_\odot h^2/Mpc^2$) \\
SigmaE\_1 & float & 2D density in elliptical rings of the 2nd radial bin & ($M_\odot h^2/Mpc^2$) \\
SigmaE\_2 & float & 2D density in elliptical rings of the 3rd radial bin & ($M_\odot h^2/Mpc^2$) \\
SigmaE\_3 & float & 2D density in elliptical rings of the 4th radial bin & ($M_\odot h^2/Mpc^2$) \\
SigmaE\_4 & float & 2D density in elliptical rings of the 5th radial bin & ($M_\odot h^2/Mpc^2$) \\
SigmaE\_5 & float & 2D density in elliptical rings of the 6th radial bin & ($M_\odot h^2/Mpc^2$) \\
SigmaE\_6 & float & 2D density in elliptical rings of the 7th radial bin & ($M_\odot h^2/Mpc^2$) \\
SigmaE\_7 & float & 2D density in elliptical rings of the 8th radial bin & ($M_\odot h^2/Mpc^2$) \\
SigmaE\_8 & float & 2D density in elliptical rings of the 9th radial bin & ($M_\odot h^2/Mpc^2$) \\
SigmaE\_9 & float & 2D density in elliptical rings of the 10th radial bin & ($M_\odot h^2/Mpc^2$) \\
SigmaE\_10 & float & 2D density in elliptical rings of the 11th radial bin & ($M_\odot h^2/Mpc^2$) \\
SigmaE\_11 & float & 2D density in elliptical rings of the 12th radial bin & ($M_\odot h^2/Mpc^2$) \\
SigmaE\_12 & float & 2D density in elliptical rings of the 13th radial bin & ($M_\odot h^2/Mpc^2$) \\
SigmaE\_13 & float & 2D density in elliptical rings of the 14th radial bin & ($M_\odot h^2/Mpc^2$) \\
SigmaE\_14 & float & 2D density in elliptical rings of the 15th radial bin & ($M_\odot h^2/Mpc^2$) \\
SigmaE\_15 & float & 2D density in elliptical rings of the 16th radial bin & ($M_\odot h^2/Mpc^2$) \\
SigmaE\_16 & float & 2D density in elliptical rings of the 17th radial bin & ($M_\odot h^2/Mpc^2$) \\
SigmaE\_17 & float & 2D density in elliptical rings of the 18th radial bin & ($M_\odot h^2/Mpc^2$) \\
SigmaE\_18 & float & 2D density in elliptical rings of the 19th radial bin & ($M_\odot h^2/Mpc^2$) \\
SigmaE\_19 & float & 2D density in elliptical rings of the 20th radial bin & ($M_\odot h^2/Mpc^2$) \\
SigmaE\_20 & float & 2D density in elliptical rings of the 21th radial bin & ($M_\odot h^2/Mpc^2$) \\
SigmaE\_21 & float & 2D density in elliptical rings of the 22th radial bin & ($M_\odot h^2/Mpc^2$) \\
SigmaE\_22 & float & 2D density in elliptical rings of the 23th radial bin & ($M_\odot h^2/Mpc^2$) \\
SigmaE\_23 & float & 2D density in elliptical rings of the 24th radial bin & ($M_\odot h^2/Mpc^2$) \\
SigmaE\_24 & float & 2D density in elliptical rings of the 25th radial bin & ($M_\odot h^2/Mpc^2$) \\
lgMDelta & float & log base 10 of the mass within a radius that encloses Delta times the critical density & \\
Delta & float & Overdensity considered to compute lgMDelta & \\
lgMNFW\_rho & float & log base 10 (M200c/$M_\odot$/h) mass from a NFW fit to the 3D density profile in spherical shells & \\
cNFW\_rho & float & c200c concentration from a NFW fit to the 3D density profile in spherical shells & \\
resNFW\_rho & float & residuals of the NFW fit to the 3D density profile in spherical shells & \\
nb\_rho & int & number of fitted bins using NFW model of the 3D density profile in spherical shells & \\
lgMNFW\_rho\_E & float & log base 10 (M200c/$M_\odot$/h) mass from a NFW fit to the 3D density profile in ellipsoidal shells & \\
cNFW\_rho\_E & float & c200c concentration from a NFW fit to the 3D density profile in ellipsoidal shells & \\
resNFW\_rho\_E & float & residuals of the NFW fit to the 3D density profile in ellipsoidal shells & \\
nb\_rho\_E & int & number of fitted bins using NFW model of the 3D density profile in ellipsoidal shells & \\
lgMNFW\_S & float & log base 10 (M200c/$M_\odot$/h) mass from a NFW fit to the 2D density profile in circular shells & \\
cNFW\_S & float & c200c concentration from a NFW fit to the 2D density profile in circular shells & \\
resNFW\_S & float & residuals of the NFW fit to the 2D density profile in circular shells & \\
nb\_S & int & number of fitted bins using NFW model of the 2D density profile in circular shells & \\
lgMNFW\_S\_E & float & log base 10 (M200c/$M_\odot$/h) mass from a NFW fit to the 2D density profile in elliptical shells & \\
cNFW\_S\_E & float & c200c concentration from a NFW fit to the 2D density profile in elliptical shells & \\
resNFW\_S\_E & float & residuals of the NFW fit to the 2D density profile in elliptical shells & \\
nb\_S\_E & int & number of fitted bins using NFW model of the 2D density profile in elliptical shells & \\
lgMEin\_rho & float & log base 10 (M200c/$M_\odot$/h) mass from a Einasto fit to the 3D density profile in spherical shells & \\
cEin\_rho & float & c200c concentration from a Einasto fit to the 3D density profile in spherical shells & \\
alpha\_rho & float & slope from a Einasto fit to the 3D density profile in spherical shells & \\
resEin\_rho & float & residuals of the Einasto fit to the 3D density profile in spherical shells & \\
lgMEin\_rho\_E & float & log base 10 (M200c/$M_\odot$/h) mass from a Einasto fit to the 3D density profile in ellipsoidal shells & \\
cEin\_rho\_E & float & c200c concentration from a Einasto fit to the 3D density profile in ellipsoidal shells & \\
alpha\_rho\_E & float & slope concentration from a Einasto fit to the 3D density profile in ellipsoidal shells & \\
resEin\_rho\_E & float & residuals of the Einasto fit to the 3D density profile in ellipsoidal shells & \\
lgMEin\_S & float & log base 10 (M200c/$M_\odot$/h) mass from a Einasto fit to the 2D density profile in circular shells & \\
cEin\_S & float & c200c concentration from a Einasto fit to the 2D density profile in circular shells & \\
alpha\_S & float & slope from a Einasto fit to the 2D density profile in circular shells & \\
resEin\_S & float & residuals of the Einasto fit to the 2D density profile in circular shells & \\
lgMEin\_S\_E & float & log base 10 (M200c/$M_\odot$/h) mass from a Einasto fit to the 2D density profile in elliptical shells & \\
cEin\_S\_E & float & c200c concentration from a Einasto fit to the 2D density profile in elliptical shells & \\
alpha\_S\_E & float & slope from a Einasto fit to the 2D density profile in elliptical shells & \\
resEin\_S\_E & float & residuals of the Einasto fit to the 2D density profile in elliptical shells & \\
\hline
    \label{tab:catalogue}
\end{longtable}


\bsp	
\label{lastpage}
\end{document}